\pdfoutput=1
\RequirePackage{ifpdf}
\ifpdf 
\documentclass[pdftex]{sigma}
\else
\documentclass{sigma}
\fi

\usepackage{csquotes}
\usepackage{mathrsfs}
\usepackage{tikz}
\usepackage{pgfplots}

\numberwithin{equation}{section}

\newtheorem{Theorem}{Theorem}[section]
\newtheorem{Corollary}[Theorem]{Corollary}
\newtheorem{Lemma}[Theorem]{Lemma}
\newtheorem{Proposition}[Theorem]{Proposition}
\newtheorem{rhp}[Theorem]{Riemann--Hilbert Problem}

 { \theoremstyle{definition}
\newtheorem{Definition}[Theorem]{Definition}
\newtheorem{Remark}[Theorem]{Remark}

\newtheorem{Example}[Theorem]{Example} }

\newcommand*{\bord}{\multicolumn{1}{c|}{}}

\makeatletter
\def\widebreve{\mathpalette\wide@breve}
\def\wide@breve#1#2{\sbox\z@{$#1#2$}%
 \mathop{\vbox{\m@th\ialign{##\crcr
\kern0.08em\brevefill#1{0.8\wd\z@}\crcr\noalign{\nointerlineskip}%
 $\hss#1#2\hss$\crcr}}}\limits}
\def\brevefill#1#2{$\m@th\sbox\tw@{$#1($}%
 \hss\resizebox{#2}{\wd\tw@}{\rotatebox[origin=c]{90}{\upshape(}}\hss$}
\makeatletter

\makeatletter

\usepackage{pict2e,picture}
\pgfkeys{/csteps/inner ysep/.initial=4pt,
 /csteps/inner xsep/.initial=4pt,
 /csteps/inner color/.initial=red,
 /csteps/outer color/.initial=blue,
}
\newsavebox\csteps@CBox
\newlength\csteps@XLength \newlength\csteps@YLength \newlength\csteps@YDepth \newlength\csteps@tmplen
\def\csteps@CircledParam#1#2{\sbox\csteps@CBox{#2}%
 \csteps@XLength=\wd\csteps@CBox\advance\csteps@XLength by\pgfkeysvalueof{/csteps/inner xsep}\relax
 \csteps@tmplen=\pgfkeysvalueof{/csteps/inner ysep}\relax
 \csteps@YDepth=\dp\csteps@CBox\advance\csteps@YDepth by 0.5\csteps@tmplen\relax
 \csteps@YLength=\ht\csteps@CBox\advance\csteps@YLength by\dp\csteps@CBox\advance\csteps@YLength by\pgfkeysvalueof{/csteps/inner ysep}\relax
 \typeout{DBG:#2\space X\space\the\csteps@XLength\space Y:\the\csteps@YLength\space D:\the\csteps@YDepth}%
 \raisebox{-#1\csteps@YDepth}{%
 \ifdim\csteps@XLength>\csteps@YLength
 \makebox[\csteps@XLength]{
 \makebox(0,\csteps@YLength){%
 \color{\pgfkeysvalueof{/csteps/outer color}}\put(0,0){\oval(\csteps@XLength,\csteps@YLength)}%
 }%
 \makebox(0,\csteps@YLength){%
 \put(-.5\wd\csteps@CBox,0){\textcolor{\pgfkeysvalueof{/csteps/inner color}}{#2}}%
 }}%
 \else
 \makebox[\csteps@YLength]{%
 \makebox(0,\csteps@YLength){%
 \color{\pgfkeysvalueof{/csteps/outer color}}\put(0,0){\circle{\csteps@YLength}}%
 }%
 \makebox(0,\csteps@YLength){%
 \put(-.5\wd\csteps@CBox,0){\textcolor{\pgfkeysvalueof{/csteps/inner color}}{#2}}%
 }}%
 \fi
 }%
}
\def\Circled#1{\csteps@CircledParam{1}{#1}}
\def\CircledTop#1{\csteps@CircledParam{0}{#1}}
\makeatother

\tikzset{/csteps/inner ysep=10pt}
\tikzset{/csteps/inner xsep=10pt}

\begin{document}
\allowdisplaybreaks

\renewcommand{\thefootnote}{}

\newcommand{\arXivNumber}{2306.14107}

\renewcommand{\PaperNumber}{076}

\FirstPageHeading

\ShortArticleName{A Riemann--Hilbert Approach to Skew-Orthogonal Polynomials of Symplectic Type}

\ArticleName{A Riemann--Hilbert Approach to Skew-Orthogonal\\ Polynomials of Symplectic Type\footnote{This paper is a~contribution to the Special Issue on Evolution Equations, Exactly Solvable Models and Random Matrices in honor of Alexander Its' 70th birthday. The~full collection is available at \href{https://www.emis.de/journals/SIGMA/Its.html}{https://www.emis.de/journals/SIGMA/Its.html}}}

\Author{Alex LITTLE}

\AuthorNameForHeading{A.~Little}

\Address{Unit\'e de Math\'ematiques Pures et Appliqu\'ees, ENS de Lyon, France}
\Email{\href{mailto:alexander.little@ens-lyon.fr}{alexander.little@ens-lyon.fr}}

\ArticleDates{Received December 27, 2023, in final form August 06, 2024; Published online August 16, 2024}

\Abstract{We present a representation of skew-orthogonal polynomials of symplectic type~($\beta=4$) in terms of a matrix Riemann--Hilbert problem, for weights of the form ${\rm e}^{-V(z)}$ where $V$ is a polynomial of even degree and positive leading coefficient. This is done by representing skew-orthogonality as a kind of multiple-orthogonality. From this, we derive a~${\beta=4}$ analogue of the Christoffel--Darboux formula. Finally, our Riemann--Hilbert representation allows us to derive a Lax pair whose compatibility condition may be viewed as a~${\beta=4}$ analogue of the Toda lattice.}


\Classification{60B20; 33C45; 30E15; 30E25}

\renewcommand{\thefootnote}{\arabic{footnote}}
\setcounter{footnote}{0}

\section{Introduction}

In this paper we develop a Riemann--Hilbert approach to the theory of skew-orthogonal polynomials of symplectic type. In order to motivate our work, let us recall the basic theory of orthogonal polynomials. One begins with a measurable function $w \colon \mathbb{R}\longrightarrow [0,+\infty]$ with finite moments to all orders, that is $\int_\mathbb{R} |x|^k w(x) {\rm d}x < +\infty$ for all $k \in \mathbb{N}$. Then one constructs a~Hermitian form
\begin{equation*}
\langle P, Q \rangle_2 = \int_\mathbb{R} P(x)Q(x) w(x) {\rm d}x
\end{equation*}
for (real) polynomials $P$ and $Q$. We would like that $\langle \cdot , \cdot \rangle_2$ be nondegenerate on the linear space of real polynomials so that we can make it into an inner product, which would be true if $w$ is continuous and has positive mass $\int_\mathbb{R} w {\rm d}x > 0$. Given such an inner product there is a unique sequence of polynomials $\{ P_n \}_{n \in \mathbb{N}}$ such that $P_n(x) = x^n + \mathcal{O}\big(x^{n-1}\big)$ and $\langle P_n , P_m \rangle_2 = 0$ for $n \neq m$, constructable by a Gram--Schmidt procedure. $w$ is referred to as the \textit{weight function} and $P_n$ is called the $n$-th monic orthogonal polynomial with respect to $w$.

The applications of orthogonal polynomials are extremely diverse, however in this work we shall be interested solely in their application to \textit{random matrix theory} (see, e.g., \cite{deiftbook}). Let $M$ be an $n\times n$ Hermitian random matrix distributed according to a probability measure
\begin{equation}\label{probmeas}
\frac{1}{Z_n} {\rm e}^{- \operatorname{tr} V(M)} {\rm d}M,
\end{equation}
where ${\rm d}M = \prod_{k=1}^n {\rm d}M_{kk} \prod_{1 \leq i < j \leq n} {\rm d} \operatorname{Re} (M_{ij}) {\rm d} \operatorname{Im} (M_{ij})$, $V$ is a real-valued analytic function on~$\mathbb{R}$ and~${Z_n\! > 0}$ is a constant of normalisation. For this measure to be normalisable, ${V(x) \to +\infty}$ as $x \to \pm \infty$ sufficiently rapidly, more precisely, $\liminf_{x \to \pm \infty} \frac{V(x)}{\log|x|} = +\infty$. This condition would be satisfied if $V$ was a polynomial of even degree and positive leading coefficient, as will be the case in this paper.

A remarkable calculation (see \cite[Chapter~5]{mehta} or \cite[Chapter~5]{loggases}) then shows that the correlation functions of the eigenvalues of such an ensemble are expressible entirely in terms of the orthogonal polynomials $P_n$ and $P_{n-1}$ and their respective norms, $\| P_n \|$ and $\| P_{n-1} \|$, with respect to the weight function $w(x) = {\rm e}^{-V(x)}$. Thus to make an asymptotic analysis of the correlation functions as $n \to \infty$ it is necessary to make an asymptotic analysis of orthogonal polynomials with respect to more or less arbitrary weight functions. For certain \enquote{classical} weight functions, this can be done by classical steepest descent techniques, however the general case required the discovery of a \textit{Riemann--Hilbert} representation of orthogonal polynomials \cite{fokas} and the development of the \textit{nonlinear steepest descent} technique \cite{10.2307/2946540}. A Riemann--Hilbert representation involves representing an object of interest in terms of the solution to a boundary value problem in the complex plane with prescribed behaviour at infinity, and can be regarded as a~generalisation of the notion of a contour integral \cite{itsRH}. This led to a proof of universality of local correlations in the bulk and at the edge of the spectrum for ensembles of type \eqref{probmeas} by means of the so-called Deift--McLaughlin--Kriecherbauer--Venakides--Zhou (DMKVZ) scheme~\mbox{\cite{deift,deift2,deift3}}, building off earlier work of Pastur--Shcherbina \cite{Pastur:1997aa} and Bleher--Its \cite{Its}.

Motivated by, amongst others, applications to quantum mechanical systems with time-re\-ver\-sal invariance (see \cite[Chapter 2]{mehta}), random matrix theorists have also been interested for a long time in studying not only Hermitian but also real symmetric and quaternion self-dual random matrices with probability measures of the form of \eqref{probmeas}. These three cases, real symmetric, complex Hermitian and quaternion self-dual, correspond to the three finite-dimensional, associative division algebras over the real numbers, $\mathbb{R}$, $\mathbb{C}$ and $\mathbb{H}$, respectively, \cite{Dyson3}. These are often labelled by the \enquote{Dyson index} $\beta=1,2,4$, respectively.

In this work, we shall be interested in the case of $\beta = 4$, the so-called symplectic case, since the probability measure is invariant under conjugation by elements of the compact symplectic group, and since we prefer to think of an $n\times n$ quaternion matrix as a $2n \times 2n$ matrix where the quaternions are represented by $2 \times 2$ blocks. For such a $2n \times 2n$ matrix, there are, with probability~1, only $n$ distinct eigenvalues since every eigenvalue occurs with multiplicity $2$ (known in the physics literature as Kramers' degeneracy). These $n$ eigenvalues have the joint PDF
\begin{align}\label{pointprocess}
\rho_n(x_1, \dots, x_n) = \frac{1}{\tilde{Z}_n} \prod_{1 \leq i < j \leq n}|x_i - x_j|^4 \prod_{k=1}^{n} {\rm e}^{-2V(x_k)},
\end{align}
where $\tilde{Z}_n $ is a normalisation constant. We shall furthermore restrict to the case of $V$ being a~real polynomial of even degree $D$ and leading coefficient $\gamma > 0$,
\[
V(x) = \gamma x^D + \mathcal{O}\big(x^{D-1}\big),\qquad x \to \infty.
\]
Here the classical approach (see \cite[Chapter 5]{mehta}, \cite[Chapter 6]{loggases}) involves finding an analogue of orthogonal polynomials and this is supplied by \textit{skew-orthogonal polynomials}. Here the inner product $\langle \cdot, \cdot \rangle_2$ is replaced by a skew-symmetric nondegenerate bilinear form $\langle \cdot, \cdot \rangle_4$, where
\begin{equation}\label{skewprod}
\langle P, Q \rangle_4 \overset{\mathrm{def}}{=} \frac{1}{2} \int_{\mathbb{R}} ( P(x) Q^\prime(x)-P^\prime(x) Q(x) ) {\rm e}^{-2V(x)} {\rm d}x
\end{equation}
for $P$ and $Q$ real polynomials. By \enquote{non-degenerate}, we mean the following. Let $\mathscr{P}_k^\mathbb{R}$ be the space of real polynomials of degree $\leq k$. Then we say $\langle \cdot, \cdot \rangle_4$ is non-degenerate if for all $n \in \mathbb{N}$, we have the implication
\begin{align*}
\forall P \in \mathscr{P}_{2n+1}^\mathbb{R} \big( \langle P, Q \rangle_4 = 0,\, \forall Q \in \mathscr{P}_{2n+1}^\mathbb{R} \Longrightarrow P = 0\big).
\end{align*}

Then a sequence of monic polynomials $\{ P_n \}_{n \in \mathbb{N}}$, $P_n(x) = x^n + \mathcal{O}\big(x^{n-1}\big)$, is said to be skew-orthogonal with respect to $\langle \cdot, \cdot \rangle_4$ if the following holds:
\begin{enumerate}\itemsep=0pt
\item[(1)] $\langle P_{2n}, P_{2m} \rangle_4 = \langle P_{2n+1}, P_{2m+1} \rangle_4 = 0$ for all $n,m \in \mathbb{N}$.
\item[(2)] $\langle P_{2n}, P_{2m+1} \rangle_4 = 0$ for all $n, m \in \mathbb{N}$, $n\neq m$.
\end{enumerate}
The existence of such a sequence is guaranteed by non-degeneracy (it can be constructed by a~procedure analogous to the Gram--Schmidt method), however these properties do not uniquely define the sequence because one can replace $P_{2n+1} \mapsto P_{2n+1} + \alpha P_{2n}$ for any constant $\alpha \in \mathbb{R}$. This degree of freedom may be fixed by demanding that the next-to-leading coefficient of $P_{2n+1}$ is equal to a specified value (it is common to choose this value to be zero however this will not be the most convenient choice for us). It can then be verified that with this extra constraint the sequence is unique.

The importance of skew-orthogonal polynomials for random matrix theory can be seen by the following. Let $h_k = \langle P_{2k}, P_{2k+1} \rangle_4$ be the \enquote{skew norm}, where by the non-degeneracy of~${\langle \cdot, \cdot \rangle_4}$ we necessarily have $h_k \neq 0$. Then (see \cite[Propositions~6.1.6 and 6.1.7]{loggases}) all eigenvalue correlation functions associated to the point process~\eqref{pointprocess} may be expressed in terms of the so-called \textit{pre-kernel},
\begin{equation}\label{prekernel}
S_n(x,y) \overset{\mathrm{def}}{=} \sum_{k=0}^{n-1} \frac{ P_{2k}(x) {\rm e}^{-V(x)} \frac{\rm d}{{\rm d}y}\big( P_{2k+1}(y) {\rm e}^{-V(y)} \big) - P_{2k+1}(x) {\rm e}^{-V(x)} \frac{\rm d}{{\rm d}y}\big( P_{2k}(y) {\rm e}^{-V(y)} \big)}{2 h_k}.
\end{equation}
Precisely, from $S_n$ one constructs a $2 \times 2$ matrix \enquote{kernel}, and then the $k$-point correlation function of the ensemble is written as a $2k \times 2k$ Pfaffian, or equivalently, quaternion determinant (see~\cite[Chapter 6]{loggases}). In particular, the 1-point correlation function is given by~\smash{$\rho_n^{(1)}(x) = S_n(x,x)$}. A~similar treatment in terms of skew-orthogonal polynomials, though with a different skew-inner product than \eqref{skewprod}, can be done in the case of real symmetric matrices ($\beta =1$).

The importance of understanding skew-orthogonal polynomials of real and symplectic type ($\beta = 1,4$) was stressed by Dyson \cite{Dyson} when he wrote:
\begin{quote}
We need to develop the theory of skew-orthogonal polynomials until it becomes a~working tool as handy as the existing theory of orthogonal polynomials.
\end{quote}
Unfortunately, this theory remains comparatively underdeveloped relative to that of orthogonal polynomials ($\beta =2$). Nevertheless let us state some known results. Adler, Horozov and van Moerbeke \cite{Adler1999ThePL} give Pfaffian formulas for the even and odd skew-orthogonal polynomials for~${\beta=1,4}$. Eynard \cite{Eynard:2001aa} derives Heine-type multiple integral formulas for even and odd skew-orthogonal polynomials for $\beta =1,4$ and he uses a non-rigorous but intriguing resolvent method to obtain~${n \to \infty}$ asymptotics for the case of a one-cut polynomial potential. These multiple-integral formulas are morally equivalent to the Pfaffian formulas since the two may be related by de Bruijn identities \cite{debruijn}. In the case of $V^\prime$ being rational, Adler and van Moerbeke \cite{adler1999pfaff} find a~natural basis of orthogonal polynomials in which to express the corresponding skew-orthogonal polynomials and Adler, Forrester, Nagao and van Moerbeke \cite{Adler:2000aa} use this to write simple expressions for the skew-orthogonal polynomials in the classical (Hermite, Laguerre and Jacobi) cases.\looseness=1

A key inspiration for the present paper is the paper of Pierce \cite{PIERCE2008230} in which a Riemann--Hilbert representation for skew-orthogonal polynomials of real type ($\beta = 1$) with a polynomial $V$ is derived; unfortunately this work has not received nearly the same attention as the corresponding problem for orthogonal polynomials because there is no known Christoffel--Darboux-type formula to express the pre-kernel in terms of the Riemann--Hilbert problem, and even if there was, the complexity of the jump matrix makes it unclear how to perform a nonlinear steepest descent analysis. Pierce's paper is similar to the present one in that it considers polynomial potentials and this allows one to represent ($\beta = 1$) skew-orthogonality as a kind of multiple-orthogonality, and this naturally leads to a Riemann--Hilbert problem.

Although skew-orthogonal polynomials on the real line have received relatively little recent attention, there has been much recent work on planar symplectic ensembles and their associated skew-orthogonal polynomials. Akemann, Ebke and Parra \cite{Akemann:2022aa} give a construction of skew-orthogonal polynomials in the plane and relate these to a corresponding set of orthogonal polynomials; while several recent works \cite{doi:10.1142/S2010326322500472,Byun_2023,byun2024scaling} derive a \enquote{generalised Christoffel--Darboux formula} for several different ensembles, which is a differential equation relating the kernel for the symplectic ensemble to that of the corresponding complex ensemble. These works are in a similar spirit to the method of Widom \cite{Widom:1999aa} in that they establish relations between the symplectic ensemble and the corresponding complex ensemble, the latter of which is often more tractable. This should be contrasted with the present work where our Riemann--Hilbert formulation does not make any direct connection with the corresponding random matrix ensemble for $\beta = 2$.

The difficulty of working with skew-orthogonal polynomials has led some to believe that they are the wrong way to approach $\beta = 1,4$ ensembles and to develop a different formalism that avoids skew-orthogonal polynomials entirely. This is the Widom formalism \cite{Widom:1999aa}, presented and developed in the excellent book of Deift and Gioev \cite{deiftuniversality}, and it is this formalism that has yielded proofs of universality of local correlations for $\beta=1,4$ \cite{https://doi.org/10.1002/cpa.20164,10.1093/imrp/rpm004,Deift:2007aa,kriecherbauer2010fluctuations,Shcherbina:2009aa,Shcherbina:2009ab,Shcherbina:2011aa,Stojanovic:2000aa}. One of the aims of this paper is to show that an approach involving skew-orthogonal polynomials can nevertheless be highly fruitful and that it is possible to treat $\beta = 4$ ensembles in a self-contained way, without making a reduction to $\beta = 2$.

Let us now state our results. Given the sequence of skew-orthogonal polynomials $\{ P_n \}_{n \in \mathbb{N}}$, we define the \enquote{dual} polynomials as \smash{$\Psi_n(x) \overset{\mathrm{def}}{=} - \frac{1}{\gamma D} {\rm e}^{V(x)} \frac{\rm d}{{\rm d}x}\big( P_{n}(x) {\rm e}^{-V(x)} \big)$}.
\begin{enumerate}\itemsep=0pt
\item[(1)] In Section \ref{rhpsection}, we show how $\Psi_{2n}$ and $\Psi_{2n+1}$ may be thought of as \enquote{multiple-orthogonal polynomials of Type II} (see Corollary~\ref{typeII}). This yields two families of Riemann--Hilbert Problems~\ref{evenrhp} and~\ref{oddrhp} ($A_n$ and $B_n$, respectively), where $\Psi_{2n}$ and $\Psi_{2n+1}$ appear as the $(1,1)$ matrix element of $A_n$ and $B_n$, respectively. $A_n$ has dimension $D+1$ and $B_n$ has dimension $D+2$. Furthermore, we find that $P_{2n}$ is the $(2,2)$ matrix element of $A_n^{-1}$, however, somewhat contrary to what one might expect, neither $P_{2n+1}$ nor any multiple of it appears as a~matrix element of $B_n^{-1}$. This seems to be related to the fact that $P_{2n}$ can be thought of as a multiple-orthogonal polynomial of Type I (see Corollary \ref{typeI}), but we are not aware of a~way to represent $P_{2n+1}$ in such a way.
\item[(2)] In Section \ref{christoffelsection}, we prove an analogue of the Christoffel--Darboux formula, namely
\begin{equation}\label{christoffel}
S_n(x,y) = - \frac{1}{4\pi {\rm i}} {\rm e}^{-V(x) - V(y)} \frac{\big( A_n^{-1}(x) A_n(y) \big)_{21}}{x-y} ,
\end{equation}
where $S_n$ is the pre-kernel \eqref{prekernel} and $A_n$ is the solution of Riemann--Hilbert Problem \ref{evenrhp}. If a~nonlinear steepest descent analysis could be carried out for~$A_n$, we would thus obtain a new proof of universality for $\beta = 4$, perhaps easier than existing proofs. It is also interesting to remark that the pre-kernel is expressible entirely in terms of the \enquote{even} Riemann--Hilbert problem $A_n$, i.e., from the point of view of random matrix theory the odd problem~$B_n$ can be ignored. This is not the first instance that an analogue of the Christoffel--Darboux formula for skew-orthogonal polynomials has appeared in the literature; in \cite{Ghosh_2006}, Ghosh derives a formula for $(x-y)S_n(x,y)$ in terms of something which, for polynomial potentials, has finite rank. However, no connection to a~Riemann--Hilbert problem was made so an asymptotic analysis of this formula was limited to special cases.
\item[(3)] In Section \ref{dynamicsection}, we show how the $A_n$ problem may be transformed to have constant jump matrices. Then allowing the potential $V$ to depend on a parameter $t \in \mathbb{R}$ by $V(z,t) = V_0(z) + zt$, we derive an $(n,t)$ Lax pair. The linear system in $n$ could be regarded as a $\beta = 4$ analogue of the \enquote{three term recurrence} which appears in the theory of orthogonal polynomials (see Remark \ref{recurrence}). The compatibility condition of this Lax pair yields a dynamical system which one could regard as a $\beta=4$ analogue of the Toda lattice (see Theorem~\ref{skewtoda}). In the $D=2$ case an exact solution can be given (see Example \ref{exactsolution}). We should remark that an analogue of the Toda lattice for $\beta = 1,4$, with connections to skew-orthogonal polynomials, already exists in the literature under the name of the \enquote{Pfaff lattice} \cite{Adler1999ThePL}. The relationship between the Pfaff lattice and our dynamical system is left as a topic for future investigation.
\end{enumerate}
As already remarked, if a nonlinear steepest descent analysis of $A_n$ could be carried out then~\eqref{christoffel} would give access to the asymptotic behaviour of many observables that were either out of reach of previous techniques or could only be approached by cumbersome methods. For example, it could be used to probe non-standard universality classes, analogous to those that have been studied for $\beta = 2$ when two cuts in the support of the equilibrium measure merge \cite{https://doi.org/10.1002/cpa.10065, https://doi.org/10.1002/cpa.20113,Claeys2005MulticriticalUR} and also in planar ensembles when one has a merging singularity \cite{kruger2024local}. Such non-standard universality classes are much less well-understood in the $\beta = 1,4$ cases as compared to $\beta = 2$.

\subsection{Notation}
\begin{enumerate}\itemsep=0pt
\item[(1)] We shall use the notation $\mathit{X}^k$ to stand for the rational function $\mathit{X}^k(x) = x^k$ for $k \in \mathbb{Z}$.
\item[(2)] For $a,b \in \mathbb{Z}$, we let $[\![a,b]\!]$ be the integer interval from $a$ to $b$, $[\![a,b]\!] = [a,b] \cap \mathbb{Z}$.
\item[(3)] For $k \in \mathbb{N}$, define the space of real polynomials of degree at most $k$ as
$\mathscr{P}_{k}^\mathbb{R}$, and let
\smash{$\mathscr{P}^\mathbb{R} = \bigcup_{k\geq 0} \mathscr{P}_{k}^\mathbb{R}$} be the space of all real polynomials. Similarly, let $\mathscr{P}^\mathbb{C}_k$ be the space of polynomials with complex coefficients of degree $\leq k$ and \smash{$\mathscr{P}^\mathbb{C} = \bigcup_{k \geq 0} \mathscr{P}^\mathbb{C}_k$}.
\item[(4)] For $m$ a positive integer and $\mathbb{F}$ a field (in our case always $\mathbb{R}$ or $\mathbb{C}$), we let $\mathcal{M}_m (\mathbb{F})$ be the space of $m \times m$ matrices with entries in $\mathbb{F}$.
\item[(5)] We use the convention that $0 \in \mathbb{N}$.
\item[(6)] We denote the 1-torus as $\mathbb{T} = \{ z \in \mathbb{C} \colon |z|=1 \}$.
\item[(7)] We let $\chi_A$ be the indicator function of the set $A$, i.e., $\chi_A(x) = 1$ if $x \in A$ and $\chi_A(x) = 0$ otherwise.
\item[(8)] Let $\Gamma \subset \mathbb{C}$ be a piecewise smooth contour in $\mathbb{C}$. Then the \textit{Cauchy transform} of the measurable function $f\colon \Gamma \longrightarrow \mathbb{C}$ is defined as
\[
C_\Gamma (f) (z) = \frac{1}{2\pi {\rm i}}\int_\Gamma \frac{f(x)}{x-z} {\rm d}x,\qquad z \in \mathbb{C}\setminus \Gamma,
\]
whenever this integral exists. In our case, we will always consider sufficiently \enquote{nice} functions $f$ and contours $\Gamma$ such that the Cauchy transform is well-defined and analytic for all $z \in \mathbb{C}\setminus \Gamma$. Furthermore, we shall always consider functions with finite moments~${\int_\Gamma |x|^k |f(x)| |{\rm d}x| < +\infty}$ for all $k\geq 0$ and for which $C_\Gamma (f)(z)$ admits an asymptotic expansion in $\frac{1}{z}$ as $z \to \infty$ in the sense of Poincar\'e. More precisely, for every $K \in \mathbb{N}$, we~have
\begin{align*}
C_\Gamma (f)(z) = - \frac{1}{2\pi {\rm i}} \sum_{k=0}^K z^{-k-1} \int_\Gamma x^k f(x) {\rm d}x +\frac{z^{-K-1}}{2\pi {\rm i} } \int_\Gamma \frac{ x^{K+1} f(x)}{x-z} {\rm d}x .
\end{align*}
If $\mathrm{dist}(z,\Gamma) \geq \epsilon > 0$, we find that $\int_\Gamma \frac{ x^{K+1} f(x)}{x-z} {\rm d}x$ is bounded. However, if $f$ is analytic we may deform the contour to achieve boundedness in a full neighbourhood of (complex) infinity. This will be the case in all the Cauchy integrals that appear in this paper. The general theory of such Cauchy integrals for functions of varying degrees of decay and regularity is discussed in notes of Deift \cite{deift2019riemannhilbert} and the book of Muskhelishvilli \cite{singular}.
\end{enumerate}

\section{Riemann--Hilbert representations}\label{rhpsection}

Before proceeding, let us demonstrate that the problem we are considering is well-posed. The following proposition is of course known to experts but we include a proof to keep the presentation complete.
\begin{Proposition} The skew-symmetric bilinear form defined by \eqref{skewprod} is non-degenerate. Furthermore, the skew-norms are all positive, i.e., $h_n = \langle P_{2n}, P_{2n+1} \rangle_4 > 0$, for all $n \geq 0$.
\end{Proposition}
\begin{proof}
By one of de Bruijn's identities \cite{debruijn}, we have
\begin{equation*}
\mathrm{pf} \big( \big\langle \mathit{X}^{i-1}, \mathit{X}^{j-1} \big\rangle_4 \big)_{i,j=1}^{2n} = \frac{1}{n!} \int_{\mathbb{R}^n} \prod_{1 \leq i < j \leq n} |x_i - x_j|^4 \prod_{k=1}^n {\rm e}^{-2V(x_k)} {\rm d} x_1 \cdots {\rm d}x_n > 0,
\end{equation*}
where $\mathrm{pf}$ is the Pfaffian. Hence the matrix $\big( \big\langle \mathit{X}^{i-1}, \mathit{X}^{j-1} \big\rangle_4 \big)_{i,j=1}^{2n}$ is invertible for all $n \geq 1$ and so defines a non-degenerate bilinear form. Furthermore, by performing row and column operations,
 \begin{align*}
0 < \mathrm{pf} \big( \big\langle \mathit{X}^{i-1}, \mathit{X}^{j-1} \big\rangle_4 \big)_{i,j=1}^{2n} = h_0 \cdots h_{n-1},\qquad \forall n \geq 1,
\end{align*}
which shows that $h_k > 0$ for all $k \geq 0$.
\end{proof}

We begin by observing that the symplectic inner product \eqref{skewprod} can be re-written as
\[
\langle P, Q \rangle_4 = \int_{\mathbb{R}} P(x) {\rm e}^{-V(x)} \frac{\rm d}{{\rm d}x}\big( Q(x) {\rm e}^{-V(x)} \big) {\rm d}x,\qquad P,Q \in \mathscr{P}^\mathbb{R}.
\]
\begin{Definition}[dual skew-orthogonal polynomials]
Given a sequence of skew-orthogonal polynomials $\{ P_n \}_{n \in \mathbb{N}}$, let us introduce the sequence of \enquote{dual} polynomials $\{ \Psi_n \}_{n \in \mathbb{N}}$, where
\[
\Psi_n(x) \overset{\mathrm{def}}{=} -\frac{1}{\gamma D} {\rm e}^{V(x)}\frac{\rm d}{{\rm d}x}\big( {\rm e}^{-V(x)} P_n(x) \big) = x^{n+D-1} + \mathcal{O}\big(x^{n+D-2}\big),\qquad x \to \infty.
\]
\end{Definition}
The monic skew-orthogonal polynomials can be reconstructed from the dual polynomials by integration,
\[
P_n(x) = \gamma D {\rm e}^{V(x)} \int_{x}^{+\infty} {\rm e}^{-V(y)} \Psi_n(y) {\rm d}y.
\]
In order to characterise $\Psi_n$, let us ask the following question. Given a polynomial $Q \in \mathscr{P}^\mathbb{C}$, under what conditions will
\begin{equation*}
{\rm e}^{V(x)} \int_{x}^{+\infty} {\rm e}^{-V(y)} Q(y) {\rm d}y
\end{equation*}
also be a polynomial? Note that this is equivalent to the question of whether $Q$ can be written $Q(x) = {\rm e}^{V(x)} \frac{\rm d}{{\rm d}x} \big( P(x) {\rm e}^{-V(x)} \big)$ for some polynomial $P$. Let us now proceed to answer this question.

\begin{Remark}\label{polyunique} Note that the map $P \mapsto {\rm e}^V \frac{\rm d}{{\rm d}x}\big(P {\rm e}^{-V}\big)$ which maps $\mathscr{P}^\mathbb{C}\to \mathscr{P}^\mathbb{C}$ is linear and injective. Thus if $Q = {\rm e}^V \frac{\rm d}{{\rm d}x}\big(P {\rm e}^{-V}\big)$ then this $P$ is unique.
\end{Remark}

Consider the contours
\begin{align}\label{contour1}
C_k \overset{\mathrm{def}}{=} {\rm e}^{2\pi {\rm i}\frac{k}{D}} [0,+\infty),\qquad k \in [\![ 0,D-1]\!].
\end{align}
Let us orient these contours so $C_0=[0,+\infty)$ has left to right orientation, whereas the remaining contours $C_1, \dots, C_{D-1}$ have orientation towards the origin. Then let
\begin{align}\label{contour2}
\Gamma_k \overset{\mathrm{def}}{=} C_0 \cup C_k,\qquad k \in [\![ 1,D-1]\!].
\end{align}
For example, when $k=\frac{D}{2}$ (recall $D$ is even and positive), $\Gamma_k = \mathbb{R}$ with standard left to right orientation. These contours are depicted in Figure \ref{contours} for the case $D=8$.

\begin{figure}
\begin{center}
\begin{tikzpicture}[scale=0.85]
		\draw[blue, thick,->] (0,0) -- (2,0);
		\draw[blue, thick] (0,0) -- (4,0);
		\draw[blue, thick] (2,0) -- node[below] {$C_0$}(4,0);
		\draw[blue,thick,->] (0,4) -- node[right] {$C_2$}(0,2);
		\draw[blue, thick] (0,0) -- (0,4);
		\draw[blue,thick] (0,0) -- (2*99/70,2*99/70);
		\draw[blue,thick,->] (2*99/70,2*99/70) -- node[right] {$C_1$} (99/70,99/70);
		\draw[blue, thick] (0,0) -- (-4,0);
		\draw[blue, thick,->] (-4,0) -- node[below] {$C_4$} (-2,0);
		\draw[blue,thick,->] (0,-4) -- node[right] {$C_6$} (0,-2);
		\draw[blue,thick] (0,0) -- (0,-4);
		\draw[blue,thick] (0,0) -- (-2*99/70,-2*99/70);
		\draw[blue, thick,->] (-2*99/70,-2*99/70) -- node[right] {$C_5$} (-99/70,-99/70);
		\draw[blue,thick] (0,0) -- (-2*99/70,2*99/70);
		\draw[blue, thick,->] (-2*99/70,2*99/70) -- node[left] {$C_3$} (-99/70,99/70);
		\draw[blue, thick] (0,0) -- (2*99/70,-2*99/70);
		\draw[blue, thick,->] (2*99/70,-2*99/70) -- node[left] {$C_7$} (99/70,-99/70);
\end{tikzpicture}
\end{center}
\caption{This figure depicts the contours $C_0, \dots, C_{D-1}$ in the case $D=8$. The arrows denote the orientation of the contours. Note the outgoing orientation of $C_0$.}\label{contours}
\end{figure}

\begin{Proposition}\label{representation} Let $Q \in \mathscr{P}^\mathbb{C}$ be a polynomial, then $Q$ can written as
\begin{equation*}
Q(x) = {\rm e}^{V(x)} \frac{\rm d}{{\rm d}x}\big({\rm e}^{-V(x)} P(x) \big) = P^\prime(x) - V^\prime(x) P(x)
\end{equation*}
for some polynomial $P \in \mathscr{P}^\mathbb{C}$ if and only if the following integrals vanish:
\begin{align}\label{integrals}
\int_{\Gamma_k} Q(x){\rm e}^{-V(x)} {\rm d}x = 0,\qquad \forall k \in [\![ 1,D-1]\!]
\end{align}
\end{Proposition}
\begin{proof}
\enquote{Only if} is trivial since for $z \in C_k \setminus \{ 0 \}$, $z^{D} > 0$, hence $P(z){\rm e}^{-V(z)}$ is rapidly decreasing, hence
\begin{align*}
\int_{\Gamma_k} Q(x){\rm e}^{-V(x)} {\rm d}z = \int_{\Gamma_k} \frac{\rm d}{{\rm d}x}\big(P(x){\rm e}^{-V(x)}\big) {\rm d}x =0,\qquad \forall k \in [\![ 1,D-1]\!].
\end{align*}
For the converse claim, suppose the integrals \eqref{integrals} all vanish. Let us then divide $D$ into ${\deg Q+1}$ with remainder, so that $\deg Q+1 = D \ell + r$, where $\ell , r \in \mathbb{N}$ and $r \leq D-1$. We will construct a~certain polynomial $R$ as follows. If $\deg Q \leq D-2$, then $R:=Q$, so we may assume ${\deg Q \geq D-1}$. We can then factorise
\begin{equation*}
Q(x) = F_1 (x) V^\prime(x) + G_1(x),
\end{equation*}
where $F_1 \in \mathscr{P}^\mathbb{C}$ has degree $\deg Q - D+1$ and $G_1 \in \mathscr{P}^\mathbb{C}$ has degree at most $D-2$. We can then repeat the procedure,
\begin{gather*}
F_1^\prime(x) = F_2(x) V^\prime(x) + G_2(x), \qquad
F_2^\prime(x)= F_3(x) V^\prime(x) + G_3(x),\qquad \ldots, \\
F_{\ell-1}^\prime(x) = F_{\ell}(x) V^\prime(x) + G_{\ell}(x).
\end{gather*}
$F_j \in \mathscr{P}^\mathbb{C}$ has degree $\deg Q-Dj+1$ and $G_j \in \mathscr{P}^\mathbb{C}$ has degree at most $D-2$ for $j=1, \dots , \ell$. In particular $F_\ell$ has degree $r$ and so the process terminates at this stage. Let
\begin{equation*}
R(x) := G_1(x) + \dots + G_\ell(x) + F_\ell^\prime(x).
\end{equation*}
$R$ has degree at most $D-2$.

By repeated integration by parts, we see that
\begin{align*}
\int_{\Gamma_k} R(x) {\rm e}^{-V(x)} {\rm d}x = 0,\qquad \forall k \in [\![ 1,D-1]\!].
\end{align*}
We now claim that $R \equiv 0$. Consider the function
\begin{equation*}
C(z) \overset{\mathrm{def}}{=} {\rm e}^{V(z)}\int_z^{+\infty} R(x) {\rm e}^{-V(x)} {\rm d}x,
\end{equation*}
where the integration path is from $z \in \mathbb{C}$ to $0$ then from $0$ to $+\infty$. $C$ is clearly an entire function. We claim that as $|z| \to +\infty$, $C(z) \to 0$, uniformly in $\operatorname{arg} z $. Hence by Liouville's theorem $C \equiv 0$, and hence $R \equiv 0$.

From our assumptions,
\begin{align*}
\int_z^{+\infty} R(x) {\rm e}^{-V(x)} {\rm d}x = \int_z^{(+\infty) \cdot {\rm e}^{2\pi {\rm i} \frac{k}{D}}} R(x) {\rm e}^{-V(x)} {\rm d}x,\qquad \forall k \in [\![ 1,D-1]\!].
\end{align*}
Now introduce the function $\varphi(z) = \tilde{V}(z)^\frac{1}{D}$ \big(where $\tilde{V}(z) = \frac{1}{\gamma}V(z)$\big) which is holomorphic and injective so long as $|z|$ is sufficiently large, i.e., on a neighbourhood of (complex) infinity. Let us choose $k \in \{0, \dots, D-1\}$, so that
\[
 2\pi \frac{k-\frac{1}{2}}{D} \leq \arg \varphi(z) \leq 2\pi \frac{k+\frac{1}{2}}{D}
\]
 (this $k$ may not be unique but this does not matter). Let us also define $\arg$ so that there is no branch cut through this sector. Finally, let us choose the contour such that $\operatorname{Re} (V(z))$ is always non-decreasing along the contour. The existence of such a contour is proven in Lemma \ref{contour}. Integrating by parts, we find
\begin{align*}
\int_z^{(+\infty) \cdot {\rm e}^{2\pi {\rm i} \frac{k}{D}}} R(x) {\rm e}^{-V(x)} {\rm d}x = \frac{R(z)}{V^\prime(z)} {\rm e}^{-V(z)}+ \int_z^{(+\infty) \cdot {\rm e}^{2\pi {\rm i} \frac{k}{D}}} \frac{\rm d}{{\rm d}x}\left( \frac{R(x)}{V^\prime(x)} \right) {\rm e}^{-V(x)} {\rm d}x.
\end{align*}
\smash{$\frac{\rm d}{{\rm d}x}\big( \frac{R(x)}{V^\prime(x)} \big)$} is a rational function which scales like $x^{-2}$ and so for $|x|$ sufficiently large there is a~constant $\kappa>0$ such that \smash{$\big| \frac{\rm d}{{\rm d}x}\big( \frac{R(x)}{V^\prime(x)} \big)\big| \leq \kappa |x|^{-2}$}. Thus
\begin{equation*}
\bigg|{\rm e}^{V(z)} \int_z^{(+\infty) \cdot {\rm e}^{2\pi {\rm i} \frac{k}{D}}} R(x) {\rm e}^{-V(x)} {\rm d}x\bigg| \leq \frac{|R(z)|}{|V^\prime(z)|}+ \kappa\int_z^{(+\infty) \cdot {\rm e}^{2\pi {\rm i} \frac{k}{D}}}|x|^{-2} |{\rm d}x| \longrightarrow 0
\end{equation*}
as $z \to \infty$. This limit holds so long as $z$ is confined to the sector \smash{$ 2\pi \frac{k-\frac{1}{2}}{D} \leq \arg \varphi(z) \leq 2\pi \frac{k+\frac{1}{2}}{D}$}, however since there are only finitely many such sectors this limit thus holds uniformly in $\operatorname{arg} \varphi(z)$ (and so uniformly in $\operatorname{arg} z$). Thus $R \equiv 0$.

Finally, this shows that ${\rm e}^{V(z)} \int_z^{+\infty} Q(x) {\rm e}^{-V(x)} {\rm d}x$ is a polynomial, since if we factorise $Q$ and integrate by parts, we~find
\begin{equation*}
{\rm e}^{V(z)} \int_z^{+\infty} Q(x) {\rm e}^{-V(x)} {\rm d}x = F_1(z) + \dots + F_\ell(z),
\end{equation*}
which is a polynomial of degree $\deg Q - D + 1$.
\end{proof}

\begin{Lemma}\label{contour} Let $\tilde{V}(w) = \frac{1}{\gamma}V(w) = w^D + \mathcal{O}\big(w^{D-1}\big)$ and let \smash{$\varphi(w) := \tilde{V}(w)^\frac{1}{D}$}. There exists a~${\varrho > 0}$ sufficiently large such that $\varphi$ is analytic and injective on
\begin{equation*}
U_\varrho := \{ z \in \mathbb{C}\colon |z| > \varrho \}
\end{equation*}
and such that the following holds. For each $z \in U_\varrho$ such that
\smash{$\frac{2\pi ( k - \frac{1}{2})}{D} \leq \arg \varphi(z) \leq \frac{2\pi ( k + \frac{1}{2})}{D}$},
there is a of contour $\tau_z$ starting at $z$ and tending to \smash{$(+\infty) \cdot {\rm e}^{2\pi {\rm i} \frac{k}{D}}$} such that $\operatorname{Re} (V(w))$ is non-decreasing along the contour and $\int_{\tau_z}|x|^{-2} |{\rm d}x| \to 0$ as $|z| \to \infty$ uniformly in $\operatorname{arg} z$.
\end{Lemma}
\begin{proof}
Let
\begin{equation*}
\tilde{\tau} (s) =
\begin{cases}
 |\varphi(z)| \exp\big({\rm i} (1-s)\arg \varphi(z) + {\rm i}s\frac{2\pi k}{D} \big), & s \in [0,1] ,\\
 (|\varphi(z)|+s-1) \exp\big({\rm i} \frac{2\pi k}{D} \big), & s \in [1,+\infty). \\
\end{cases}
\end{equation*}
Then we choose our contour to be the preimage of $\tilde{\tau}$, that is $\tau= \varphi^{-1} \circ \tilde{\tau}$. From this, we see that $V(\tau(s)) = \gamma \tilde{\tau}(s)^D$. A short calculation then shows that for $s \in (0,1)$
\begin{equation*}
\frac{\rm d}{{\rm d}s} \operatorname{Re} (V(\tau(s))) = \gamma |\varphi(z)|^D D \left( \arg \varphi(z) - \frac{2\pi k}{D} \right) \sin\left( D (1-s) \left( \arg \varphi(z) - \frac{2\pi k}{D} \right) \right) \geq 0.
\end{equation*}
It is also easy to show for $s \in (1,+\infty)$, $\frac{\rm d}{{\rm d}s} \bigl(\operatorname{Re} \tilde{\tau}(s)^D\bigr) \geq 0$. Furthermore, $\varphi(w) = w (1+o(1))$ and~${\varphi^\prime(w) = 1+o(1)}$ uniformly on $U_\varrho$ as $\varrho \to +\infty$, and so $\int_\tau |x|^{-2} |{\rm d}x| \to 0$ as $|z| \to \infty$ uniformly in $\operatorname{arg} z$.
\end{proof}

A central result in the theory of orthogonal polynomials is that their zeros are simple and lie in the support of the weight function. The following proposition gives a partial extension of this to dual skew-orthogonal polynomials.
\begin{Proposition}
$\Psi_{2n}$ has at least $2n+1$ real zeros $($not counting multiplicity$)$, of which at least $2n+2-\frac{D}{2}$ must be simple. $\Psi_{2n+1}$ has at least $2n$ real zeros $($not counting multiplicity$)$, of which at least $2n - \frac{D}{2}$ must be simple.
\end{Proposition}
\begin{proof}
Let $ x_1, \dots, x_\ell$ be the collection of real zeros of $\Psi_{2n}$ with odd multiplicity, i.e., exactly those points on $\mathbb{R}$ at which $\Psi_{2n}$ changes sign, where we do not repeat according to multiplicity. Then $\Psi_{2n}(x)(x-x_1)\cdots (x-x_\ell)$ has the same sign everywhere and vanishes only on a finite set, hence
\begin{equation*}
\int_\mathbb{R} \Psi_{2n}(x)(x-x_1)\cdots (x-x_\ell) {\rm e}^{-2V(x)} {\rm d}x \neq 0 .
\end{equation*}
However, if $\ell \leq 2n$ then the above integral vanishes, thus $\ell \geq 2n+1$. Furthermore, it is easily seen that at least $2n+2-\frac{D}{2}$ of these must be simple. A similar argument works for $\Psi_{2n+1}$.
\end{proof}

We now arrive at a pair of Riemann--Hilbert problems, for the even and odd dual skew-orthogonal polynomials, respectively. In what follows, it is more convenient to ensure the uniqueness of the sequence by fixing the next-to-leading coefficient of~$\Psi_{2n+1}$ to be zero rather than~$P_{2n+1}$. We begin with the \enquote{even problem}, so called because the $(1,1)$ matrix element of the solution will turn out to be $\Psi_{2n}$.

\begin{rhp}[even problem]\label{evenrhp} Let $V$ be a real polynomial of even degree~${D \geq 2}$ and positive leading coefficient, $\Gamma_j$ be defined by \eqref{contour1} and \eqref{contour2}, and
\begin{equation}\label{gamma}
\Gamma = \bigcup_{j=1}^{D-1} \Gamma_j
\end{equation}
with the orientations of $\Gamma_j$ inherited by $\Gamma$ $($see, e.g., Figure {\rm\ref{contours}}$)$. We look for a $(D+1) \times (D+1)$ matrix valued function
\begin{equation*}
A_{n} \colon\ \mathbb{C}\setminus \Gamma \longrightarrow \mathcal{M}_{D+1}(\mathbb{C})
\end{equation*}
such that
\begin{enumerate}\itemsep=0pt
\item[$(1)$] $A_{n}$ is analytic on $\mathbb{C}\setminus \Gamma$ $($that is, its matrix elements are analytic on $\mathbb{C}\setminus \Gamma)$.
\item[$(2)$] $A_{n}$ has continuous boundary values up to $\Gamma \setminus \{ 0 \}$. That is, given $x \in \Gamma \setminus \{ 0 \}$, as $\mathbb{C}\setminus \Gamma \ni z \to x$ from either the $+$ $($left$)$ side or $-$ $($right$)$ side non-tangentially, the limit of $A_{n}(z)$ exists. These limits are denoted
\begin{align*}
A_{n}^+ \colon\ \Gamma \setminus \{ 0 \} \longrightarrow \mathcal{M}_{D+1}(\mathbb{C}),\qquad
A_{n}^- \colon\ \Gamma \setminus \{ 0 \} \longrightarrow \mathcal{M}_{D+1}(\mathbb{C})
\end{align*}
and are continuous functions. \enquote{Left} and \enquote{right} mean relative to the orientation of the contour. The boundary values are related by the \enquote{jump matrix}
\begin{align*}
A_{n}^{+}(x) = A_{n}^{-}(x)\left(
\begin{array}{ccccccccccccccccccc}
1 & {\rm e}^{-2V} \chi_\mathbb{R} & {\rm e}^{-V} \chi_{\Gamma_1} & \dots & {\rm e}^{-V} \chi_{\Gamma_{D-1}} \\
& 1 \\
& & 1 \\
& & & \ddots \\
& & & & 1 \\
\end{array} \right), \qquad x \in \Gamma \setminus \{ 0 \}.
\end{align*}
\item[$(3)$] $A_{n}$ is bounded in a neighbourhood of $0$.
\item[$(4)$] We have the asymptotic normalisation
\begin{align*}
A_{n}(z) = \big(\mathbb{I}+ \mathcal{O}\big(z^{-1}\big)\big) \left(\begin{matrix}
z^{2n+D-1} \\
& z^{-2n} \\
& & z^{-1} \\
& & & \ddots \\
& & & & z^{-1}
\end{matrix} \right),\qquad z \to \infty.
\end{align*}
\end{enumerate}
\end{rhp}
\begin{Lemma}\label{uniqueness} If a solution to Riemann--Hilbert Problem {\rm \ref{evenrhp}} exists, then it is unique.
\end{Lemma}
\begin{proof}
This method is standard but we repeat it for completeness. We note that the jump matrix has determinant $1$ so $\det A_{n}(z) $ has no jump across $\Gamma \setminus \{ 0 \}$ and takes continuous boundary values, so by Morera's theorem is analytic everywhere except possibly at $0$. However, the requirement of boundedness in a neighbourhood of $0$ implies that the singularity at $0$ is removable. Hence $z \mapsto \det A_{n}(z)$ is entire. Furthermore, we have the normalisation
$\det A_{n}(z) = 1 + \mathcal{O}\big(z^{-1}\big)$,
 as $z \to \infty$ and so $\det A_{n}(z) \equiv 1$.
Then let \smash{$\tilde{A}_{n}$} be another solution. Then the function \smash{$z \mapsto \tilde{A}_{n}(z) A_{n}^{-1}(z)$}
has no jump across $\Gamma \setminus \{0 \}$ and obtains continuous boundary values, so by Morera's theorem is analytic everywhere except possibly at $0$. Again by the requirement of boundedness of \smash{$\tilde{A}_n$} and $A_n$ in a neighbourhood of $0$ and that $\det A_n(z) \equiv 1$, we see that the singularity at $0$ is removable. Hence \smash{$z \mapsto \tilde{A}_{n}(z) A_{n}^{-1}(z)$} is actually entire. Finally, since
\begin{equation*}
\tilde{A}_{n}(z) A_{n}^{-1}(z) = \mathbb{I}+\mathcal{O}\big(z^{-1}\big),
\end{equation*}
we find that $\tilde{A}_{n}(z) A_{n}^{-1}(z) \equiv \mathbb{I}$.
\end{proof}

The jump condition implies that $(A_{n})_{11}$ is an entire function which scales like $z^{2n+D-1}$ as $z \to \infty$. It is therefore a monic polynomial of degree $2n+D-1$ by Liouville's theorem,
\begin{equation*}
(A_{n})_{11} := Q \in \mathscr{P}_{2n+D-1}^\mathbb{C}.
\end{equation*}
The Plemelj formula then implies that the first row of $A_{n}$ is
\begin{equation*}
\big( Q , C_{\mathbb{R}}\big({\rm e}^{-2V} Q\big) , C_{\Gamma_1}\big({\rm e}^{-V} Q\big) , \dots , C_{\Gamma_{D-1}}\big({\rm e}^{-V} Q\big)\big).
\end{equation*}
 From the asymptotic condition
 \begin{align*}
 C_{\Gamma_j}\big({\rm e}^{-V} Q\big)(z) = \mathcal{O}\big(z^{-2}\big),\qquad \forall j \in [\![ 1,D-1]\!],
 \end{align*}
we see, by Proposition \ref{representation}, that there exists a unique $P \in \mathscr{P}^\mathbb{C}$ of degree exactly $2n$ such that~${Q = {\rm e}^{V} \frac{\rm d}{{\rm d}x}\big(P {\rm e}^{-V}\big)}$. Finally, the normalisation
\begin{align*}
C_{\mathbb{R}}\big({\rm e}^{-2V} Q\big)(z) = \mathcal{O}\big(z^{-2n-1}\big) ,\qquad z \to \infty
\end{align*}
implies that $\langle \mathit{X}^k , P\rangle_4 =0 $ for $k \in [\![0, , 2n-1]\!]$, and so $P$ is skew-orthogonal. Thus $Q = \Psi_{2n}$.

For the second row, we see that $(A_{n})_{21} := \tilde{Q}$ is a polynomial of degree at most $2n+D-2$. Following the same procedure as above, we see that the second row is
\begin{equation*}
\big( \tilde{Q} , C_{\mathbb{R}}\big({\rm e}^{-2V} \tilde{Q}\big) , C_{\Gamma_1}\big({\rm e}^{-V} \tilde{Q}\big) , \dots , C_{\Gamma_{D-1}}\big({\rm e}^{-V} \tilde{Q}\big)
\big),
\end{equation*}
where now $\tilde{Q} = - \frac{2\pi {\rm i} \gamma D}{h_{n-2}} \Psi_{2n-2}$. For the remaining rows, we see that $(A_n)_{j+2,1} := R^{(j)}_n \in \mathscr{P}^{\mathbb{C}}_{2n+D-2}$ must be a polynomial of degree at most $2n+D-2$. To show there exists a unique solution to Riemann--Hilbert Problem~\ref{evenrhp}, let us introduce the following linear functionals:
\begin{align*}
\alpha_j \colon\ \mathscr{P}^{\mathbb{C}}_{2n+D-2} \longrightarrow \mathbb{C},\qquad
\beta_j \colon\ \mathscr{P}^{\mathbb{C}}_{2n+D-2} \longrightarrow \mathbb{C},
\end{align*}
where
\begin{align}
&\alpha_j(P) \overset{\mathrm{def}}{=} \int_\mathbb{R} {\rm e}^{-2V(x)} x^j P(x) {\rm d}x,\qquad j \in [\![ 0, 2n-1 ]\!],\label{alpha} \\
&\beta_j(P) \overset{\mathrm{def}}{=} -\frac{1}{2\pi {\rm i}} \int_{\Gamma_j} {\rm e}^{-V(x)} P(x) {\rm d}x,\qquad j \in [\![ 1, D-1]\!]. \label{beta}
\end{align}
\begin{Lemma}\label{linindep}The linear functionals $\alpha_0, \dots, \alpha_{2n-1}, \beta_1, \dots, \beta_{D-1}$ form a linearly independent set $\big($and thus span $\big(\mathscr{P}_{2n+D-2}^{\mathbb{C}}\big)^\ast\big)$.
\end{Lemma}

\begin{proof}
To show this, we need only show that
\begin{equation*}
\bigcap_{j=0}^{2n-1} \ker \alpha_j \cap \bigcap_{k=1}^{D-1} \ker \beta_k = 0.
\end{equation*}
Thus let \smash{$Q \in \mathscr{P}^\mathbb{C}_{2n+D-2}$} be such that $\alpha_0(Q) = \dots = \alpha_{2n-1}(Q)= \beta_1(Q) = \dots = \beta_{D-1}(Q)=0$. Thus by Proposition \ref{representation}, $Q = {\rm e}^V \frac{\rm d}{{\rm d}x}\big( P {\rm e}^{-V}\big)$ for some \smash{$P \in \mathscr{P}^\mathbb{C}_{2n-1}$}. Furthermore, $\smash{\big\langle \mathit{X}^0, P \big\rangle_4 =} \dots = \big\langle \mathit{X}^{2n-1}, P \big\rangle_4 = 0$. Hence $P$ is skew-orthogonal to all of \smash{$\mathscr{P}^\mathbb{C}_{2n-1}$}. Thus by the non-degeneracy of the inner product $\langle \cdot,\cdot \rangle_4$, $P \equiv 0$, and hence $Q \equiv 0$.
\end{proof}

Let \smash{$R^{(j)}_n \in \mathscr{P}^\mathbb{C}_{2n+D-2}$} be the unique polynomial such that
\smash{$
\alpha_0\big(R^{(j)}_n\big) = \dots = \alpha_{2n-1}\big(R^{(j)}_n\big) = 0
$}
and
\begin{align*}
\beta_k\big(R^{(j)}_n\big) = \delta_{kj},\qquad \forall k \in [\![ 1, D-1 ]\!].
\end{align*}
That such a polynomial exists and is unique follows from the linear independence of these functionals.

\begin{Theorem}
The unique solution of the \textit{even} Riemann--Hilbert Problem~{\rm \ref{evenrhp}} is
\begin{gather*}
A_{n}
=\resizebox{.87\hsize}{!}{$\left(\begin{matrix}
\Psi_{2n} & C_\mathbb{R}\big({\rm e}^{-2V}\!\Psi_{2n}\big) & C_{\Gamma_1}\big({\rm e}^{-V}\!\Psi_{2n}\big) &\cdots & C_{\Gamma_{D-1}}\big({\rm e}^{-V}\!\Psi_{2n}\big) \\[1mm]
\!\!- \frac{2\pi {\rm i} \gamma D}{h_{n-1}}\Psi_{2n-2}\!\!\!\! & - \frac{2\pi {\rm i} \gamma D}{h_{n-1}} C_\mathbb{R}\big({\rm e}^{-2V}\!\Psi_{2n-2}\big)\!\!\!\! & - \frac{2\pi {\rm i} \gamma D}{h_{n-1}} C_{\Gamma_1}\big({\rm e}^{-V}\!\Psi_{2n-2}\big)\!\!\!\! & \cdots & \!\!\!\!\!- \frac{2\pi {\rm i} \gamma D}{h_{n-1}} C_{\Gamma_{D-1}}\big({\rm e}^{-V}\!\Psi_{2n-2}\big) \!\!\\[1mm]
R^{(1)}_n & C_\mathbb{R}\big({\rm e}^{-2V}\!R^{(1)}_n\big) & C_{\Gamma_1}\big({\rm e}^{-V}\!R^{(1)}_n\big) & \cdots & C_{\Gamma_{D-1}}\big({\rm e}^{-V}R^{(1)}_n\big) \\
\vdots \\
R^{(D-1)}_n & C_\mathbb{R}\big({\rm e}^{-2V}\!R^{(D-1)}_n \big) & C_{\Gamma_1}\big({\rm e}^{-V}\!R^{(D-1)}_n \big) & \cdots & C_{\Gamma_{D-1}}\big({\rm e}^{-V}R^{(D-1)}_n \big) \\
\end{matrix}\right)$}.
\end{gather*}
\end{Theorem}

\begin{Lemma}[symmetry for even case]\label{EvenSymmetry} Let $P \in \mathcal{M}_{D-1}(\mathbb{R})$ be the permutation matrix on $\mathbb{R}^{D-1}$ for which $P e_k = e_{D-k}$ for $k = 1, \dots, D-1$. Here $e_k = (0, \dots, 0, 1, 0, \dots, 0)^\mathsf{T}$, where the $1$ lies in the $k$-th entry. Let
\begin{equation*}
Q := \mathrm{diag}(-1,1) \oplus P = \left( \begin{matrix} -1 & 0 & \bord & \\
0 & 1 & \bord &\\ \hline
& & \bord & P\end{matrix} \right) \in \mathcal{M}_{D+1}(\mathbb{R}).
\end{equation*}
Note that $Q = Q^\mathsf{T} = Q^{-1}$ is an orthogonal matrix. Then we have the relation $
A_{n}(z) = Q \overline{A_{n}(\overline{z}) } Q$.
\end{Lemma}
\begin{proof}
Let \smash{$\tilde{A}_{n}(z) = Q \overline{A_{n}(\overline{z})} Q$}. We show that \smash{$\tilde{A}_{n}$} solves the even Riemann--Hilbert problem and thus by uniqueness (see Lemma \ref{uniqueness}) \smash{$\tilde{A}_{n} = A_{n}$}. Clearly, \smash{$\tilde{A}_{n}$} is analytic on $\mathbb{C}\setminus \Gamma$ and it is easy to see that
\begin{align*}
\tilde{A}_{n}(z) = \big(\mathbb{I}+ \mathcal{O}\big(z^{-1}\big)\big) \left(\begin{matrix}
z^{2n+D-1} \\
& z^{-2n} \\
& & z^{-1} \\
& & & \ddots \\
& & & & z^{-1}
\end{matrix} \right),\qquad z \to \infty.
\end{align*}
Furthermore, $\tilde{A}_{n}$ is clearly bounded at $0$ since $A_{n}$ is.

Because $V$ is a polynomial with real coefficients, $\overline{V(\overline{x})} = V(x)$. Furthermore, $\chi_\mathbb{R}(\overline{x}) = \chi_\mathbb{R}(x)$ and $\chi_{\Gamma_k}(\overline{x}) = \chi_{\Gamma_{D-k}}(x)$. Now for $x \in \Gamma \setminus \{ 0 \}$, we have
\begin{align*}
\big(\tilde{A}_{n} \big)_+(x) &= Q \overline{( A_{n})_- (\overline{x})} Q \\
&= Q \overline{( A_{n})_+ (\overline{x})} \left(\begin{array}{ccccccccccccccccccc}
1 & -{\rm e}^{-2V} \chi_\mathbb{R} & -{\rm e}^{-V} \chi_{\Gamma_{D-1}} & \dots & -{\rm e}^{-V} \chi_{\Gamma_{1}} \\
& 1 \\
& & 1 \\
& & & \ddots \\
& & & & 1 \\
\end{array} \right) Q \\
&=\big(\tilde{A}_{n} \big)_-(x)\left(\begin{array}{ccccccccccccccccccc}
1 & {\rm e}^{-2V} \chi_\mathbb{R} & {\rm e}^{-V} \chi_{\Gamma_{1}} & \dots & {\rm e}^{-V} \chi_{\Gamma_{D-1}} \\
& 1 \\
& & 1 \\
& & & \ddots \\
& & & & 1 \\
\end{array} \right).
\end{align*}
Thus $\tilde{A}_{n} = A_{n}$.
\end{proof}

\begin{Remark} Our method assumes $V$ is a polynomial, however there are interesting cases, e.g., Jacobi weights, that do not fit into this class. How might our analysis be extended? The natural way would be to redefine $\Psi_n$ and \smash{$R_n^{(j)}$} so that these are no longer polynomials. For example, if $V^\prime$ was rational with simple poles at $\lambda_1, \dots, \lambda_k$, then $\Psi_n$ would also be rational with simple poles, with the property that $\mathrm{Res}_{z = \lambda_i} \Psi_{n}(z) {\rm e}^{-V(z)} = 0$ for all $i=1,\dots, k$ and similarly for the auxiliary functions \smash{$R^{(j)}_n$}. Investigation of this is left as a topic for future research.
\end{Remark}
We now move onto the \enquote{odd problem}, so called because it will later turn out that the $(1,1)$ matrix element of the solution is $\Psi_{2n+1}$.
\begin{rhp}[odd problem]\label{oddrhp}
Let $V$ be a real polynomial of even degree~${D \geq 2}$ and positive leading coefficient, and let
\begin{equation*}
\Sigma = \mathbb{T} \cup \bigcup_{j=1}^{D-1} \Gamma_j,
\end{equation*}
where $\Gamma_j$ have the specified orientations and $\mathbb{T}$ is positively oriented $($see, e.g., Figure {\rm\ref{morecontours}}$)$. Let \smash{$\omega = {\rm e}^{\frac{2\pi {\rm i} }{D}}$} be a primitive root of unity and let
\begin{equation*}
S = \big\{ 0, 1, \omega, \omega^2, \dots, \omega^{D-1} \big\}
\end{equation*}
be the points of self-intersection of $\Sigma$. We look for a $(D+2) \times (D+2)$ matrix valued function
\begin{equation*}
B_{n} \colon\ \mathbb{C}\setminus \Sigma \longrightarrow \mathcal{M}_{D+2}(\mathbb{C})
\end{equation*}
such that
\begin{enumerate}\itemsep=0pt
\item[$(1)$] $B_{n}$ is analytic on $\mathbb{C}\setminus \Sigma$ $($that is, its matrix elements are analytic on $\mathbb{C}\setminus \Sigma)$.
\item[$(2)$] $B_{n}$ has continuous boundary values up to $\Sigma \setminus S$. That is, given $x \in \Sigma \setminus S$, as $\mathbb{C}\setminus \Sigma \ni z \to x$ from either the $+$ $($left$)$ side or $-$ $($right$)$ side non-tangentially, the limit of $B_{n}(z)$ exists. These limits are denoted
\begin{align*}
B_{n}^+\colon\ \Sigma \setminus S \longrightarrow \mathcal{M}_{D+2}(\mathbb{C}),\qquad
B_{n}^-\colon\ \Sigma \setminus S \longrightarrow \mathcal{M}_{D+2}(\mathbb{C})
\end{align*}
and are continuous functions. \enquote{Left} and \enquote{right} mean relative to the orientation of the contour. The boundary values are related by the \enquote{jump matrix}
\begin{gather*}
B_{n}^{+}(x) = B_{n}^{-}(x)\left(
\begin{array}{ccccccccccccccccccc}
1 & {\rm e}^{-2V} \chi_\mathbb{R} & {\rm e}^{-V} \chi_{\Gamma_1} & \dots & {\rm e}^{-V} \chi_{\Gamma_{D-1}} & \mathit{X}^{-2n-D} \chi_\mathbb{T} \\
& 1 \\
& & 1 \\
& & & \ddots \\
& & & & 1 \\
& & & & & 1 \\
\end{array} \right), \\
 x \in \Sigma \setminus S.
\end{gather*}
\item[$(3)$] $B_{n}$ is bounded in a neighbourhood of $S$.
\item[$(4)$] We have the asymptotic normalisation
\begin{align*}
B_{n}(z) = \big(\mathbb{I}+ \mathcal{O}\big(z^{-1}\big)\big) \left(\begin{matrix}
z^{2n+D} \\
& z^{-2n} \\
& & z^{-1} \\
& & & \ddots \\
& & & & z^{-1}
\end{matrix} \right),\qquad z \to \infty.
\end{align*}
\end{enumerate}
\end{rhp}

\begin{figure}
\label{morecontours}
\begin{center}
\begin{tikzpicture}[scale=0.85]
\draw[blue, thick,->] (0,0) -- (1,0);
		\draw[blue, thick,->] (1,0) -- (3,0);
		\draw[blue, thick] (3,0) -- node[below] {$C_0$}(4,0);
		\draw[blue, thick] (0,0) -- (4,0);
		\draw[blue,thick,->] (0,4) -- node[right] {$C_2$}(0,3);
		\draw[blue,thick,->] (0,3) -- (0,1);
		\draw[blue, thick] (0,0) -- (0,4);
		\draw[blue,thick] (0,0) -- (99/70,99/70);
		
		\draw[blue,thick,->] (2*99/70,2*99/70) -- node[right] {$C_1$} (1.5*99/70,1.5*99/70);
		
		\draw[blue,thick,->] (2*99/70,2*99/70) -- (0.5*99/70,0.5*99/70);
		\draw[blue, thick] (0,0) -- (-4,0);
		\draw[blue, thick,->] (-4,0) -- node[below] {$C_4$} (-3,0);
		\draw[blue, thick,->] (-3,0) -- (-1,0);
		
		\draw[blue, thick,->] (-2*99/70,2*99/70) -- node[left] {$C_3$} (-1.5*99/70,1.5*99/70);
		\draw[blue, thick,->] (-2*99/70,2*99/70) -- (-0.5*99/70,0.5*99/70);
		\draw[blue, thick] (0,0) -- (-2*99/70,2*99/70);
		
		\draw[blue,thick,->] (-2*99/70,-2*99/70) -- node[right] {$C_5$} (-1.5*99/70,-1.5*99/70);
		\draw[blue,thick] (-2*99/70,-2*99/70) -- (0,0);
		\draw[blue,thick,->] (-2*99/70,-2*99/70) -- (-0.5*99/70,-0.5*99/70);
		
		\draw[blue,thick,->] (0,-4) -- node[right] {$C_6$}(0,-3);
		\draw[blue,thick,->] (0,-3) -- (0,-1);
		\draw[blue, thick] (0,0) -- (0,-4);

		\draw[blue, thick,->] (2*99/70,-2*99/70) -- node[left] {$C_7$} (1.5*99/70,-1.5*99/70);
		\draw[blue, thick,->] (2*99/70,-2*99/70) -- (0.5*99/70,-0.5*99/70);
		\draw[blue, thick] (0,0) -- (2*99/70,-2*99/70);
		
		\draw[blue, thick,->](2*0.92387,-2*0.38268) arc (-22.5:22.5:2);
		\draw[blue, thick,->](2*0.92387,2*0.382683) arc (22.5:45+22.5:2);
		\draw[blue, thick,->](2*0.92387,2*0.382683) arc (22.5:45+22.5:2);
		\draw[blue, thick,->](2*0.92387,2*0.382683) arc (22.5:45+22.5:2);
		\draw[blue, thick,->](2*0.382683,2*0.923879) arc (45+22.5:45+45+22.5:2);
		\draw[blue, thick,->](-2*0.38268,2*0.92387) arc (45+45+22.5:45+45+45+22.5:2);
		\draw[blue, thick,->](-2*0.92387,2*0.3826834) arc (45+45+45+22.5:45+45+45+45+22.5:2);
		\draw[blue, thick,->](-2*0.923879,-2*0.382683) arc (45+45+45+45+22.5:45+45+45+45+45+22.5:2);
		\draw[blue, thick,->](-2*0.382683,-2*0.92387) arc (45+45+45+45+45+22.5:45+45+45+45+45+45+22.5:2);
		\draw[blue, thick,->](2*0.3826834,-2*0.923879) arc (45+45+45+45+45+45+22.5:45+45+45+45+45+45+45+22.5:2);
		\draw[blue, thick,->] (0,0) circle (2);
		\node[blue] at (2*0.92387+0.35,2*0.382683) {$\mathbb{T}$};
\end{tikzpicture}
\end{center}
\caption{The contours of the odd Riemann--Hilbert problem, here shown for the case $D=8$. Orientations of the contours are indicated by arrows.}
\end{figure}

The method of solution of this Riemann--Hilbert problem is very similar to the even case. The asymptotic condition implies that $(B_{n})_{11} = Q$ is a monic polynomial of degree $2n+D$. From the jump condition and the Plemelj formula we find that the first row is
\begin{equation*}
\big( Q , C_{\mathbb{R}}\big({\rm e}^{-2V} Q\big) , C_{\Gamma_1}\big({\rm e}^{-V} Q\big) , \dots , C_{\Gamma_{D-1}}\big({\rm e}^{-V} Q\big) , C_{\mathbb{T}}\big( \mathit{X}^{-2n-D} Q\big)
\big).
\end{equation*}
From the normalisation \smash{$C_{\Gamma_{j}}\big({\rm e}^{-V} Q\big)(z) = \mathcal{O}\big(z^{-2}\big)$} for all $j \in [\![1, D-1 ]\!]$, we find that there exists a polynomial $P$ of degree $2n+1$ such that $Q = {\rm e}^V \frac{\rm d}{{\rm d}x}\big( P {\rm e}^{-V}\big)$.
\begin{Remark} The scaling $C_{\mathbb{T}}\big( \mathit{X}^{-2n-D} Q\big)(z) = \mathcal{O}\big(z^{-2}\big)$ implies that
\begin{align*}
Q(x) = x^{2n+D} + \mathcal{O}\big(x^{2n+D-2}\big),\qquad z \to \infty .
\end{align*}
This fixes the next-to-leading coefficient of $P$ \big(where $Q = {\rm e}^V \frac{\rm d}{{\rm d}x}\big( P {\rm e}^{-V} \big)$\big) though not necessarily to $0$. This explains the remark made in the introduction that fixing the next-to-leading coefficient of $P$ to $0$ is not always the most convenient choice. The addition of the contour $\mathbb{T}$ could also be done for the Riemann--Hilbert problem of Pierce \cite{PIERCE2008230} in order to guarantee a unique solution.
\end{Remark}

Finally, the scaling $C_{\mathbb{R}}\big({\rm e}^{-2V} Q\big)(z) \!=\! \mathcal{O}\big(z^{-2n-1}\big)$ implies that ${\big\langle \mathit{X}^{0}, P \big\rangle_4\! = \!\cdots\! = \! \big\langle \mathit{X}^{2n-1}, P \big\rangle_4\! =\! 0}$, so $P$ is indeed skew-orthogonal. Thus $Q = \Psi_{2n+1}$.

In the same manner as for the even case, we find that the second row is
\begin{equation*}
\big( \tilde{Q} , C_{\mathbb{R}}\big({\rm e}^{-2V} \tilde{Q}\big) , C_{\Gamma_1}\big({\rm e}^{-V} \tilde{Q}\big) , \dots , C_{\Gamma_{D-1}}\big({\rm e}^{-V} \tilde{Q}\big) , C_{\mathbb{T}}\big( \mathit{X}^{-2n-D} \tilde{Q}\big) \big),
\end{equation*}
 where $\tilde{Q} = -\frac{2\pi {\rm i} \gamma D}{h_{n-1}} \Psi_{2n-2}$. Indeed, the remaining rows except the final one are the same as in the even case. To show a unique solution in all remaining rows, let us introduce the linear functional~${\beta_D \colon \mathscr{P}^\mathbb{C}_{2n+D-2} \longrightarrow \mathbb{C}}$ defined by
\[
\beta_D(P) \overset{\mathrm{def}}{=}- \frac{1}{2\pi {\rm i}} \oint_\mathbb{T} P(z) z^{-2n-D} {\rm d}z.
\]
Now let $\alpha_0, \dots, \alpha_{2n-1}$ and $\beta_1, \dots, \beta_{D-1}$ act on $\mathscr{P}^\mathbb{C}_{2n+D-1}$ according to the formulas given in~\eqref{alpha} and \eqref{beta}. That the 2nd to $(D+2)$-th rows have a unique solution follows from the lemma below.
\begin{Lemma} Consider the linear functionals $\alpha_0, \dots, \alpha_{2n-1}$ and $\beta_1, \dots, \beta_{D}$ viewed as acting on $\mathscr{P}^\mathbb{C}_{2n+D-1}$ according to \eqref{alpha} and \eqref{beta}. Then
\begin{equation*}
\bigcap_{j=0}^{2n-1} \ker \alpha_j \cap \bigcap_{k=1}^{D} \ker \beta_k = 0.
\end{equation*}
\end{Lemma}
\begin{proof}
If $\beta_D (Q) = 0$, then $Q \in \mathscr{P}^\mathbb{C}_{2n+D-2}$. The proof then reduces to that of Lemma \ref{linindep}.
\end{proof}

\begin{Theorem}
The unique solution of the \textit{odd} Riemann--Hilbert Problem~{\rm\ref{oddrhp}} is then
\begin{gather*}
B_{n} =
\resizebox{.87\hsize}{!}{$\left(\begin{matrix}
\Psi_{2n+1} & C_\mathbb{R}\big({\rm e}^{-2V}\!\Psi_{2n+1}\big) & C_{\Gamma_1}\big({\rm e}^{-V}\!\Psi_{2n+1}\big) & \cdots & C_{\mathbb{T}}\big(\mathit{X}^{-2n-D}\Psi_{2n+1}\big) \\[1mm]
\!\!- \frac{2\pi {\rm i} \gamma D}{h_{n-1}}\Psi_{2n-2}\!\!\!\! & - \frac{2\pi {\rm i} \gamma D}{h_{n-1}} C_\mathbb{R}\big({\rm e}^{-2V}\!\Psi_{2n-2}\big)\!\!\!\! & - \frac{2\pi {\rm i} \gamma D}{h_{n-1}} C_{\Gamma_1}\big({\rm e}^{-V}\!\Psi_{2n-2}\big)\!\!\!\! & \cdots &\!\!\!\!\! - \frac{2\pi {\rm i} \gamma D}{h_{n-1}} C_{\mathbb{T}}\big(\mathit{X}^{-2n-D}\Psi_{2n-2}\big)\!\! \\[1mm]
R^{(1)}_n & C_\mathbb{R}\big({\rm e}^{-2V}\!R^{(1)}_n\big) & C_{\Gamma_1}\big({\rm e}^{-V}\!R^{(1)}_n\big) & \cdots & C_{\mathbb{T}}\big(\mathit{X}^{-2n-D}R^{(1)}_n\big) \\
\vdots \\
R^{(D-1)}_n & C_\mathbb{R}\big({\rm e}^{-2V}\!R^{(D-1)}_n\big) & C_{\Gamma_1}\big({\rm e}^{-V}\!R^{(D-1)}_n\big) & \cdots & C_{\mathbb{T}}\big(\mathit{X}^{-2n-D}R^{(D-1)}_n\big) \\[1mm]
R^{(D)}_n & C_\mathbb{R}\big({\rm e}^{-2V}\!R^{(D)}_n\big) & C_{\Gamma_1}\big({\rm e}^{-V}\!R^{(D)}_n\big) & \cdots & C_{\mathbb{T}}\big(\mathit{X}^{-2n-D} R^{(D)}_n\big)
\end{matrix}\right)$}.
\end{gather*}
\end{Theorem}
Note that these polynomials \smash{$R^{(1)}_n, \dots, R^{(D-1)}_n$} are the same polynomials that appear in the even problem, so we have not made a confusion of notation. Note also that the integrals in the final column can be computed explicitly.

\begin{Proposition} \smash{$R^{(D)}_n = - \Psi_{2n}$}.
\end{Proposition}
\begin{proof}
From the normalisation, we see that \smash{$R^{(D)}_n$} is a polynomial of degree at most $2n+D-1$, and from \smash{$C_\mathbb{T}\big( \mathit{X}^{-2n-D} R^{(D)}_n\big)(z) = z^{-1}+\mathcal{O}\big(z^{-2}\big)$}, we find that \smash{$R^{(D)}_n(x) = - Q(x)$}, where $Q$ is a~monic polynomial of degree $2n+D-1$. From the fact that
\[
\beta_1\big(R^{(D)}_n\big) = \dots = \beta_{D-1}\big(R^{(D)}_n\big) = 0,
\] we find that \smash{$Q = {\rm e}^V \frac{\rm d}{{\rm d}x}\big( P {\rm e}^{-V}\big)$} for some polynomial $P$ of degree exactly $2n$. Finally, the requirement that
\[
\alpha_0\big(R^{(D)}_n\big) = \dots = \alpha_{2n-1}\big(R^{(D)}_n\big) = 0
\]
 implies that $P$ is skew-orthogonal to all of~$\mathscr{P}_{2n-1}^\mathbb{C}$.
\end{proof}

From this, one might think the $B_n$ problem is strictly superior to the $A_n$ problem, since the former encodes both $\Psi_{2n+1}$ and $\Psi_{2n}$, however our Christoffel--Darboux type formula \eqref{CD} implies that from the point of view of random matrix theory it is completely sufficient to study~$A_n$.
\begin{Lemma} \smash{$ 2n \leq \deg R_n^{(j)} \leq 2n+D-2$} for $j \in [\![1, D-1]\!]$.
\end{Lemma}
\begin{proof}
Suppose by way of contradiction that $\smash{\deg R_n^{(j)}} \leq 2n-1$. Then by the conditions \smash{$\alpha_0 \big(R_n^{(j)}\big) = \dots = \alpha_{2n-1}\big( R_n^{(j)} \big) = 0$}, we find that \smash{$R_n^{(j)} \equiv 0$}. But this contradicts \smash{$\beta_j \big(R_n^{(j)}\big) = 1$}.
\end{proof}

\begin{Remark}\label{zero} For the case $n=0$, the polynomials \smash{$R_0^{(1)} , \dots, R_0^{(D-1)}$} can be constructed as follows. Let $M$ be the matrix
\begin{equation*}
M = (M_{ij})_{i,j=1}^{D-1} = \left( \int_{\Gamma_i} x^{j-1} {\rm e}^{-V(x)} {\rm d}x \right)_{i,j=1}^{D-1} ,
\end{equation*}
then we have
\begin{align*}
R^{(j)}_0(x) = - 2\pi {\rm i} \sum_{k=1}^{D-1} \big(M^{-1}\big)_{kj} x^{k-1},\qquad j \in [\![1,D-1]\!].
\end{align*}
\end{Remark}

\begin{Lemma}[symmetry for odd problem]\label{OddSymmetry} Let $P \in \mathcal{M}_{D-1}(\mathbb{R})$ be defined as in Lemma {\rm\ref{EvenSymmetry}} and let
\begin{equation*}
Q := \mathrm{diag}(-1,1) \oplus P \oplus (-1) = \left( \begin{array}{ccccccc} -1 & 0 & \bord & \\
0 & 1 & \bord & &\\ \cline{1-5}
& & \bord & P & \bord & \\\cline{4-6}
& & & & \bord & -1 \end{array} \right) \in \mathcal{M}_{D+2}(\mathbb{R}) .
\end{equation*}
Note that $Q = Q^\mathsf{T}= Q^{-1}$ is an orthogonal matrix. Then we have the relation $B_{n}(z) = Q \overline{B_{n}(\overline{z}) } Q$.
\end{Lemma}
\begin{proof}
Morally, the same as for the even case.
\end{proof}

Finally, our results allow us to characterise the \enquote{dual} skew-orthogonal polynomials $\Psi_{2n}$ and auxiliary polynomials \smash{$R^{(j)}_n$} in terms of the corresponding sequence of \textit{orthogonal} polynomials. This is morally the same construction as in Remark \ref{zero}.

\begin{Corollary}[skew-orthogonal polynomials in terms of orthogonal polynomials]\label{orthog} Let $\{\! H_{\!n}\! \}_{\!n \in \mathbb{N}}$ be the sequence of monic orthogonal polynomials corresponding to the weight ${\rm e}^{-2V}$, where $V(x) = \gamma x^D+ \mathcal{O}\big(x^{D-1}\big)$ is a real polynomial of even degree $D \geq 2$ and $\gamma > 0$. That is,
\[H_n(x) = x^n + \mathcal{O}\big(x^{n-1}\big)\]
and
\begin{align*}
\int_\mathbb{R}H_n(x) H_m(x) {\rm e}^{-2V(x)} {\rm d}x = 0,\qquad \forall n,m \in \mathbb{N},\qquad n \neq m.
\end{align*}
Let $M$ be the matrix
\begin{equation*}
M = ( M_{jk})_{j,k=1}^{D-1} = \bigg( \int_{\Gamma_j} H_{2n+k-1}(z) {\rm e}^{-V(z)} {\rm d}z \bigg)_{j,k=1}^{D-1},
\end{equation*}
and let $\lambda_n$ be the next-to-leading coefficient of $H_{n+D}$, i.e.,
\begin{equation*}
H_{n+D}(x) = x^{n+D}+\lambda_n x^{n+D-1} + \mathcal{O}\big(x^{n+D-2}\big).
\end{equation*}
Then the corresponding \enquote{dual} skew-orthogonal polynomials and auxiliary polynomials may be expressed
\begin{gather*}
\Psi_{2n} = H_{2n+D-1} - \sum_{k,j=1}^{D-1} \bigg[ \big(M^{-1}\big)_{kj}\int_{\Gamma_j} H_{2n+D-1}(z){\rm e}^{-V(z)} {\rm d}z \bigg] H_{2n+k-1}, \\
\Psi_{2n+1} = H_{2n+D} - \lambda_n H_{2n+D-1} \\
\phantom{\Psi_{2n+1} =}{}- \sum_{k,j=1}^{D-1} \bigg[ \big(M^{-1}\big)_{kj}\int_{\Gamma_j} (H_{2n+D}(z) - \lambda_n H_{2n+D-1}(z)){\rm e}^{-V(z)} {\rm d}z \bigg] H_{2n+k-1} ,\\
R^{(j)}_n = - 2\pi {\rm i} \sum_{k=1}^{D-1} \big(M^{-1}\big)_{kj} H_{2n+k-1}.
\end{gather*}
\end{Corollary}
Indeed, these $\{ H_k \}_{k \in \mathbb{N}}$ are exactly the \enquote{natural basis} identified by Adler and van Moerbeke~\cite{adler1999pfaff}. A similar construction to Corollary \ref{orthog} in which orthogonal and skew-orthogonal polynomials were related was considered recently in the planar case by \cite{Akemann:2022aa}, however we note that in Corollary \ref{orthog} the relation is really between orthogonal polynomials and \textit{dual} skew-orthogonal polynomials.
\begin{Example}
In the case $D=2$, for the potential $V(z) = \frac{1}{2}z^2$, we can write the corresponding orthogonal polynomials explicitly as
\[
H_n(z) = \frac{(-1)^n}{2^n} {\rm e}^{z^2} \frac{{\rm d}^n}{{\rm d}z^n} {\rm e}^{-z^2}.
\]
Then using \smash{$\int_\mathbb{R}H_{2n}(x) {\rm e}^{-\frac{1}{2}x^2} {\rm d}x = \sqrt{2} \left( n - \frac{1}{2} \right)! $}, one finds that
\begin{gather*}
\begin{split}
&\Psi_{2n} = H_{2n+1}, \qquad
\Psi_{2n+1} = H_{2n+2} - \left( n +\frac{1}{2} \right) H_{2n}, \qquad P_{2n} = n! \sum_{k=0}^n \frac{1}{k!} H_{2k}, \\
& R^{(1)}_n = -\frac{2\pi {\rm i} }{ \sqrt{2} \left( n - \frac{1}{2} \right)!} H_{2n},
\qquad
P_{2n+1} = H_{2n+1}.
\end{split}
\end{gather*}
These formulas for $P_{2n}$ and $P_{2n+1}$ already appeared in the paper of Adler, Forrester, Nagao and van Moerbeke \cite{Adler:2000aa}.
\end{Example}
What we have done in the preceding discussion amounts to representing $\Psi_{2n}$ and $\Psi_{2n+1}$ as \enquote{multiple-orthogonal polynomials of Type II} (see, e.g., \cite{Martinez_Finkelshtein_2016} for a definition of this term). Let us summarise this with the following corollary.
\begin{Corollary}[dual skew-orthogonal polynomials as Type II multiple-orthogonal polynomials]\label{typeII} Consider the following problem. Find a monic polynomial $Q$ of degree exactly $2n+D-1$ such that
\begin{gather*}
\int_\mathbb{R} x^k Q(x) {\rm e}^{-2V(x)} {\rm d}x = 0,\qquad\forall k \in [\![ 0, 2n-1]\!], \\
\int_{\Gamma_j} Q(x) {\rm e}^{-V(x)} {\rm d}x = 0,\qquad \forall j \in [\![1, D-1 ]\!].
\end{gather*}
Then this problem is uniquely solved by $Q = \Psi_{2n}$. Similarly, if we are asked to find a monic polynomial $Q$ of degree exactly $2n+D$ such that
\begin{gather*}
\int_\mathbb{R} x^k Q(x) {\rm e}^{-2V(x)} {\rm d}x = 0,\qquad \forall k \in [\![ 0, 2n-1]\!], \\
\int_{\Gamma_j} Q(x) {\rm e}^{-V(x)} {\rm d}x = 0,\qquad \forall j \in [\![1, D-1 ]\!], \\
\int_{\mathbb{T}} Q(x) x^{-2n-D} {\rm d}x = 0,
\end{gather*}
then this problem is uniquely solved by $Q = \Psi_{2n+1}$. We should note that our case differs from usual multiple-orthogonality because our \enquote{measures} are not all supported on $\mathbb{R}$.
\end{Corollary}
Following the standard theory of multiple-orthogonal polynomials \cite{VanAssche2001}, it is natural to consider the Riemann--Hilbert problems satisfied by \smash{$\widehat{A_n} := A_n^{- \mathsf{T}}$} and \smash{$\widehat{B_n} := B_n^{- \mathsf{T}}$} (inverse transpose).
\begin{rhp}[dual even problem]\label{dualeven} By virtue of the fact that $\det A_n \equiv 1$, \smash{$\widehat{A_n} \overset{\mathrm{def}}{=} A_n^{-\mathsf{T}}$} satisfies the following Riemann--Hilbert problem:
\begin{enumerate}\itemsep=0pt
\item[$(1)$] \smash{$\widehat{A_{n}}$} is analytic on $\mathbb{C}\setminus \Gamma$ $($that is, its matrix elements are analytic on $\mathbb{C}\setminus \Gamma)$.
\item[$(2)$] \smash{$\widehat{A_{n}}$} has continuous boundary values up to $\Gamma \setminus \{ 0 \}$. That is, given $x \in \Gamma \setminus \{ 0 \}$, as $\mathbb{C}\setminus \Gamma \ni z \to x$ from either the $+$ $($left$)$ side or $-$ $($right$)$ side non-tangentially, the limit of \smash{$\widehat{A_{n}}(z)$} exists. These limits are denoted
\begin{align*}
\widehat{A_{n}}^+ \colon\ \Gamma \setminus \{ 0 \} \longrightarrow \mathcal{M}_{D+1}(\mathbb{C}), \qquad
\widehat{A_{n}}^- \colon\ \Gamma \setminus \{ 0 \} \longrightarrow \mathcal{M}_{D+1}(\mathbb{C})
\end{align*}
and are continuous functions. \enquote{Left} and \enquote{right} mean relative to the orientation of the contour. The boundary values are related by the \enquote{jump matrix},
\begin{align*}
\widehat{A_{n}}^{+}(x) = \widehat{A_{n}}^{-}(x)\left(
\begin{array}{ccccccccccccccccccc}
1 & & & \\
- {\rm e}^{-2V} \chi_\mathbb{R} & 1 \\
- {\rm e}^{-V} \chi_{\Gamma_1} & & & 1 \\
\vdots & & & & & \ddots \\
- {\rm e}^{-V} \chi_{\Gamma_{D-1}} & & & & & & & 1 \\
\end{array} \right), \qquad x \in \Gamma \setminus \{ 0 \}.
\end{align*}
\item[$(3)$] \smash{$\widehat{A_{n}}$} is bounded in a neighbourhood of $0$.
\item[$(4)$] We have the asymptotic normalisation
\begin{align*}
\widehat{A_{n}}(z) = \big(\mathbb{I}+ \mathcal{O}\big(z^{-1}\big)\big) \left(\begin{matrix}
z^{-2n-D+1} \\
& z^{2n} \\
& & z \\
& & & \ddots \\
& & & & z
\end{matrix} \right),\qquad z \to \infty.
\end{align*}
\end{enumerate}
\end{rhp}
In what follows we shall only be interested in computing the second column of \smash{$\widehat{A_n}$} (the motivation for this is that this column will appear in the upcoming Christoffel--Darboux-type formula), though it is an easy exercise for the reader to compute the remaining columns. By virtue of the jump condition, the 2nd to $(D+1)$-th columns all have no jump across $\Gamma \setminus \{ 0 \}$ and by boundedness around $0$ we see that the singularity at $0$ is removable. Hence the 2nd to~$(D+1)$-th columns are entire functions, indeed polynomials by the asymptotic condition,
\begin{gather*}
\big( \widehat{A_n} \big)_{i2} (z) = z^{2n} \delta_{i,2}+\mathcal{O}\big(z^{2n-1}\big),\qquad i \in [\![1, D+1]\!], \\
\big( \widehat{A_n} \big)_{ij} (z) = z \delta_{ij} + c_{ij},\qquad i \in [\![1, D+1]\!] \qquad\text{and}\qquad j \in [\![3,D+1]\!],
\end{gather*}
where the $c_{ij}$ are constants. The jump condition on the first column may be solved by Plemelj's formula
\begin{gather*}
\big( \widehat{A_n} \big)_{i1} = - C_\mathbb{R}\big( \big(\widehat{A_n}\big)_{i2} {\rm e}^{-2V} \big) - \sum_{j=3}^{D+1} c_{ij} C_{\Gamma_{j-2}}\big( {\rm e}^{-V}\big) - \chi_{i \geq 3} C_{\Gamma_{i-2}}\big( \mathit{X} {\rm e}^{-V} \big),\qquad i \in \![\![ 1, D+1]\!].
\end{gather*}
The asymptotic condition $\big( \widehat{A_n} \big)_{11}(z) = z^{-2n-D+1}\big(1 + \mathcal{O}\big(z^{-1}\big)\big)$ yields the conditions
\begin{align*}
\int_\mathbb{R} \big(\widehat{A_n}\big)_{12}(x) x^k {\rm e}^{-2V(x)} {\rm d}x + \sum_{j=3}^{D+1} c_{1j} \int_{\Gamma_{j-2}} x^k {\rm e}^{-V(x)} {\rm d}x = 0,\qquad \forall k \in [\![0, 2n+D-3 ]\!] .
\end{align*}
By taking linear combinations of these equations, we can replace $x^k$ with any polynomial of degree $\leq 2n+D-3$; in particular, we can replace it with $\Psi_0, \dots, \Psi_{2n-2}$. This implies that~$\big(\widehat{A_n}\big)_{12}$ is a~scalar multiple of $P_{2n-2}$. This multiple is fixed by
\begin{align*}
\int_\mathbb{R} \big(\widehat{A_n}\big)_{12}(x) \Psi_{2n-1}(x) {\rm e}^{-2V(x)} {\rm d}x = 2 \pi {\rm i},
\end{align*}
and so
\begin{align}\label{12Ahat}
\big(\widehat{A_n}\big)_{12} = - \frac{2\pi {\rm i} \gamma D}{h_{n-1}} P_{2n-2}.
\end{align}
By a similar argument, we see that
\begin{align}\label{22Ahat}
\big(\widehat{A_n}\big)_{22} = P_{2n}.
\end{align}
The remaining entries in the second column are polynomials of degree at most $2n-1$. These satisfy the conditions
\begin{gather*}
\int_\mathbb{R} \big(\widehat{A_n}\big)_{i2}(x) \Psi_k(x) {\rm e}^{-2V(x)} {\rm d}x + \int_{\Gamma_{i-2}} x \Psi_k(x) {\rm e}^{-V(x)} {\rm d}x = 0,\\ \forall k \in [\![0 , 2n-1]\!]\qquad \text{and}\qquad i \geq 3.
\end{gather*}
Expanding this in the basis $P_0, \dots, P_{2n-1}$, we find
\begin{gather}
\big(\widehat{A_{n}}\big)_{i2} =\gamma D \sum_{k=0}^{n-1} h_k^{-1} \left( P_{2k} \int_{\Gamma_{i-2}} x \Psi_{2k +1}(x) {\rm e}^{-V(x)} {\rm d}x - P_{2k+1} \int_{\Gamma_{i-2}} x \Psi_{2k}(x) {\rm e}^{-V(x)} {\rm d}x \right) \nonumber \\
 i \in [\![ 3,D+1]\!]. \label{i2Ahat}
\end{gather}
What the above calculation demonstrates is that $P_{2n}$ may be thought of as a Type I multiple-orthogonal polynomial (again see \cite{Martinez_Finkelshtein_2016} for a definition of this term). Let us formalise this with the following corollary.
\begin{Corollary}[even skew-orthogonal polynomials as Type I multiple-orthogonal polynomials]\label{typeI} Consider the following problem. Find a monic polynomial $P$ of degree exactly $2n$ and a~collection of constants $c_1, \dots, c_{D-1}$ such that
\begin{align*}
\int_{\mathbb{R}} x^k P(x) {\rm e}^{-2V(x)} {\rm d}x + \sum_{j=1}^{D-1} c_j \int_{\Gamma_j} x^k {\rm e}^{-V(x)} {\rm d}x = 0,\qquad \forall k \in [\![0 , 2n+D-2 ]\!].
\end{align*}
Then this problem has a unique solution $(P, c_1, \dots, c_{D-1})$, where $P=P_{2n}$.
\end{Corollary}
It is curious that Types I and~II multiple orthogonality both make their appearance here (a~similar situation, in which both Types~I and~II multiple orthogonality appear in the study of the same ensemble, was observed recently in certain planar ensembles \cite{Berezin_2023,Lee_2019}). On the other hand, it is not clear if it is possible to think of~$P_{2n+1}$ as a Type~I multiple-orthogonal polynomial and, as we shall soon see, $P_{2n+1}$ (or any multiple of it) does not appear as a matrix element of~$B_n^{-1}$. It is possible to characterise $P_{2n+1}$ by a \textit{combination} of both Types~I and~II orthogonality conditions, as follows. We look for a~monic polynomial $P$ of degree $2n+1$ and a~collection of constants $c_1, \dots, c_{D-1}$ such that
\begin{align*}
\int_\mathbb{R} P(x) x^k {\rm e}^{-2V(x)} {\rm d}x + \sum_{j=1}^{D-1} c_j \int_{\Gamma_j} x^k {\rm e}^{-V(x)} {\rm d}x = 0,\qquad \forall k \in [\![ 0, 2n+D-2]\!]
\end{align*}
and
\begin{equation*}
 \int_\mathbb{T} P(x) {\rm e}^{-V(x)} \frac{\rm d}{{\rm d}x}\big( x^{-2n-D} {\rm e}^{V(x)} \big) {\rm d}x = 0.
\end{equation*}
$(P, c_1, \dots, c_{D-1})=(P_{2n+1}, \ast, \dots, \ast)$ is the unique solution of this problem. However, this construction seems somewhat ugly and we shall not investigate the implications of this here.
\begin{rhp}[dual odd problem] By virtue of the fact that $\det B_n \equiv 1$, \smash{$\widehat{B_n}\overset{\mathrm{def}}{=}B_n^{-\mathsf{T}}$} satisfies the following Riemann--Hilbert problem:
\begin{enumerate}\itemsep=0pt
\item[$(1)$] \smash{$\widehat{B_{n}}$} is analytic on $\mathbb{C}\setminus \Sigma$ $($that is, its matrix elements are analytic on $\mathbb{C}\setminus \Sigma)$.
\item[$(2)$] \smash{$\widehat{B_{n}}$} has continuous boundary values up to $\Sigma \setminus S$. That is, given $x \in \Sigma \setminus S$, as $\mathbb{C}\setminus \Sigma \ni z \to x$ from either the $+$ $($left$)$ side or $-$ $($right$)$ side non-tangentially, the limit of \smash{$\widehat{B_{n}}(z)$} exists. These limits are denoted
\begin{align*}
\widehat{B_{n}}^+ \colon\ \Sigma \setminus S \longrightarrow \mathcal{M}_{D+2}(\mathbb{C}), \qquad
\widehat{B_{n}}^- \colon\ \Sigma \setminus S \longrightarrow \mathcal{M}_{D+2}(\mathbb{C})
\end{align*}
and are continuous functions. \enquote{Left} and \enquote{right} mean relative to the orientation of the contour. The boundary values are related by the \enquote{jump matrix}
\begin{align*}
\widehat{B_{n}}^{+}(x) = \widehat{B_{n}}^{-}(x)\left(
\begin{array}{ccccccccccccccccccc}
1 & & & \\
- {\rm e}^{-2V} \chi_\mathbb{R} & 1 \\
- {\rm e}^{-V} \chi_{\Gamma_1} & & & 1 \\
\vdots & & & & & \ddots \\
- {\rm e}^{-V} \chi_{\Gamma_{D-1}} & & & & & & & 1 \\
- \mathit{X}^{-2n-D} \chi_{\mathbb{T}} & & & & & & & & 1 \\
\end{array} \right),\qquad x \in \Sigma \setminus S.
\end{align*}
\item[$(3)$] \smash{$\widehat{B_{n}}$} is bounded in a neighbourhood of $S$.
\item[$(4)$] We have the asymptotic normalisation
\begin{align*}
\widehat{B_{n}}(z) = \big(\mathbb{I}+ \mathcal{O}\big(z^{-1}\big)\big) \left(\begin{matrix}
z^{-2n-D} \\
& z^{2n} \\
& & z \\
& & & \ddots \\
& & & & z
\end{matrix} \right), \qquad z \to \infty.
\end{align*}
\end{enumerate}
\end{rhp}
Here again we restrict ourselves to computing the second column. We shall simply state the result since the calculation is practically the same as in the case of $\widehat{A_n}$,
\begin{gather*}
\big(\widehat{B_n}\big)_{12} = 0 ,\qquad
\big(\widehat{B_n}\big)_{22} = P_{2n},\qquad
\big(\widehat{B_n}\big)_{i2} = \big(\widehat{A_n}\big)_{i2},\qquad i \in [\![3, D+1]\!] ,\\
\big(\widehat{B_n}\big)_{D+2,2} = \frac{2\pi {\rm i} \gamma D}{h_{n-1}} P_{2n-2}.
\end{gather*}

\section{Christoffel--Darboux type formula}\label{christoffelsection}

In this section, we prove the following theorem.
\begin{Theorem}[Christoffel--Darboux type formula]
Let the pre-kernel $S_n$ be defined by \eqref{prekernel}. Then
\begin{equation}\label{CD}
S_n(x,y) = -\frac{1}{4\pi {\rm i}} {\rm e}^{-V(x)-V(y)} \frac{\big( A_n^{-1}(x) A_n(y) \big)_{21}}{x-y},
\end{equation}
where $A_n$ is the unique solution of Riemann--Hilbert Problem {\rm\ref{evenrhp}}. Note that we do not specify whether we take the $+$ or $-$ boundary value of $\big( A_n^{-1}(x) A_n(y) \big)_{21}$ since this will turn out to be an entire function.
\end{Theorem}
\begin{Remark} We call this a \enquote{Christoffel--Darboux type} formula because an analogous formula appears in the theory of orthogonal polynomials ($\beta = 2$), i.e., if one takes $A_n$ to be the ${2\times 2}$ Fokas--Its--Kitaev Riemann--Hilbert problem \cite{fokas}, then \eqref{CD} yields the classical Christoffel--Darboux formula, up to a~factor of~$\frac{1}{2}$.
\end{Remark}

\begin{Remark} We could, if we like, have expressed the Christoffel--Darboux type formula~\eqref{CD} in terms of the solution of the odd Riemann--Hilbert Problem~\ref{oddrhp}, since we see from our solution to $B_n$ and \smash{$\widehat{B_n}$} that $\big( A_n^{-1}(x) A_n(y) \big)_{21} = \big( B_n^{-1}(x) B_n(y) \big)_{21}$.
\end{Remark}

To prove this result, we need to derive a variety of relations. First note that the polynomial \smash{$R_k^{(j)}- R_n^{(j)}$} for $k=0, \dots, n-1$ is in the kernel of $\beta_j$ for all $j$ and has degree at most $2n+D-2$. It is therefore in the span of $\Psi_0 , \dots, \Psi_{2n-1}$. Thus write
\[
R_k^{(j)}- R_n^{(j)} = \sum_{i=0}^{n-1} \eta_{ik}^{(j)} \Psi_{2i}+\sum_{i=0}^{n-1} \xi_{ik}^{(j)} \Psi_{2i+1}.
\]
 We then have
\begin{align*}
\eta_{ik}^{(j)} = \frac{\gamma D}{h_i}\int_\mathbb{R} R_k^{(j)} P_{2i+1} {\rm e}^{-2V} {\rm d}x ,\qquad
\xi_{ik}^{(j)} = - \frac{\gamma D}{h_i} \int_\mathbb{R} R_k^{(j)} P_{2i} {\rm e}^{-2V} {\rm d}x.
\end{align*}
Note that for $i \leq k-1$, we have \smash{$\eta_{ik}^{(j)} = \xi_{ik}^{(j)} = 0$}. We may thus write
\begin{align*}
R_k^{(j)} = R_n^{(j)}+ \sum_{i=k}^{n-1} \eta_{ik}^{(j)} \Psi_{2i}+\sum_{i=k}^{n-1} \xi_{ik}^{(j)} \Psi_{2i+1}.
\end{align*}

Next consider $\mathit{X} \Psi_{2k+1} - \Psi_{2k+2}$. This has degree at most $2n+D$ and is orthogonal to \smash{$\mathscr{P}_{2k-2}^\mathbb{C}$}. Then \smash{$\mathit{X} \Psi_{2k+1}\!- \Psi_{2k+2}\! - \sum_{j=1}^{D-1}\! \beta_j ( \mathit{X} \Psi_{2k+1}) R_k^{(j)}$} has degree at most $2n+D$ and is in the kernel of $\beta_j$ for all $j=1, \dots, D-1$. It is thus a linear combination of $\Psi_{2k-2}$, $\Psi_{2k}$ and $\Psi_{2k+1}$. We thus find
\begin{align*}
\mathit{X} \Psi_{2k+1} ={}& \Psi_{2k+2} + \sum_{j=1}^{D-1} \beta_j ( \mathit{X} \Psi_{2k+1}) R_k^{(j)} + a_k \Psi_{2k+1} + b_k \Psi_{2k} + c_k \Psi_{2k-2} \\
={}& \Psi_{2k+2} + a_k \Psi_{2k+1} + b_k \Psi_{2k} + c_k \Psi_{2k-2} + \sum_{j=1}^{D-1} \beta_j ( \mathit{X} \Psi_{2k+1}) R_n^{(j)}\\
& + \sum_{j=1}^{D-1} \beta_j ( \mathit{X} \Psi_{2k+1}) \sum_{i=k}^{n-1} \eta_{ik}^{(j)} \Psi_{2i} + \sum_{j=1}^{D-1} \beta_j ( \mathit{X} \Psi_{2k+1}) \sum_{i=k}^{n-1} \xi_{ik}^{(j)} \Psi_{2i+1} .
\end{align*}
These coefficients $a_k$, $b_k$ and $c_k$ are unique as can be seen from the formulas
\begin{align*}
a_k &=
 - \frac{\gamma D}{h_k} \int_\mathbb{R} x P_{2k} \Psi_{2k+1} {\rm e}^{-2V} {\rm d}x -\sum_{j=1}^{D-1} \beta_j ( \mathit{X} \Psi_{2k+1}) \xi^{(j)}_{kk} , \\
b_k &= \frac{\gamma D}{h_k} \int_\mathbb{R} x \Psi_{2k+1} P_{2k+1} {\rm e}^{-2V} {\rm d}x - \sum_{j=1}^{D-1}\beta_j( \mathit{X} \Psi_{2k+1}) \eta^{(j)}_{kk} , \qquad
c_k = - \frac{h_k}{h_{k-1}} .
\end{align*}
Similarly, $\mathit{X} \Psi_{2k}- \Psi_{2k+1}$ has degree at most $2n+D-1$ and is orthogonal to \smash{$ \mathscr{P}_{2k-1}^\mathbb{C}$}. Thus \smash{$\mathit{X}\Psi_{2k} - \Psi_{2k+1} - \sum_{j=1}^{D-1} \beta_j(\mathit{X}\Psi_{2k} ) R_k^{(j)}$} is in the kernel of all $\beta_j$ and is orthogonal to \smash{$\mathscr{P}_{2k-1}^\mathbb{C}$}. Thus it must be proportional to $\Psi_{2k}$. This gives
\begin{align}
\mathit{X}\Psi_{2k} ={}& \Psi_{2k+1}+ \sum_{j=1}^{D-1} \beta_j(\mathit{X}\Psi_{2k} ) R_k^{(j)} + \tilde{a}_k \Psi_{2k} \nonumber\\
={}& \Psi_{2k+1} + \tilde{a}_k \Psi_{2k} + \sum_{j=1}^{D-1} \beta_j(\mathit{X}\Psi_{2k} ) R_n^{(j)} + \sum_{j=1}^{D-1} \beta_j(\mathit{X}\Psi_{2k} ) \sum_{i=k}^{n-1} \eta_{ik}^{(j)} \Psi_{2i}\nonumber \\
& + \sum_{j=1}^{D-1} \beta_j(\mathit{X}\Psi_{2k} ) \sum_{i=k}^{n-1} \xi_{ik}^{(j)} \Psi_{2i+1}.\label{formula}
\end{align}
As before, $\tilde{a}_k$ is unique and may be written
\begin{align*}
\tilde{a}_k &= \frac{\gamma D}{h_k}\int_\mathbb{R} x \Psi_{2k} P_{2k+1} {\rm e}^{-2V} {\rm d}x - \sum_{j=1}^{D-1} \beta_j(\mathit{X}\Psi_{2k} ) \eta^{(j)}_{kk}.
\end{align*}
Equation \eqref{formula} also implies the intriguing identity
\begin{align}\label{intriguing}
2 = - \sum_{j=1}^{D-1} \beta_j(\mathit{X}\Psi_{2k} ) \xi_{kk}^{(j)}.
\end{align}
These formulas imply \enquote{dual} formulas for $P_{2k}$ and $P_{2k+1}$. Let us begin with the even case. $\mathit{X} P_{2k}$ is monic of degree $2k+1$. Thus consider $\mathit{X} P_{2k} - P_{2k+1}$ which has degree at most $2k$. Thus we may expand
\[
\mathit{X} P_{2k} - P_{2k+1} = \sum_{m= 0}^{k} \mu_{mk} P_{2m} + \sum_{m = 0}^{k-1} \lambda_{mk} P_{2m+1}.
\]
We then find
\begin{align*}
\mu_{mk} &= -\frac{\gamma D}{h_m} \int_\mathbb{R} x P_{2k} \Psi_{2m+1} {\rm e}^{-2V} {\rm d}x = \frac{h_k}{h_m} \sum_{j=1}^{D-1} \beta_j(\mathit{X} \Psi_{2m+1}) \xi_{km}^{(j)} + a_k \delta_{km}, \\
\lambda_{mk} & = \frac{\gamma D}{h_m}\int_\mathbb{R} x P_{2k} \Psi_{2m} {\rm e}^{-2V} {\rm d}x = - \frac{h_k}{h_m} \sum_{j=1}^{D-1} \beta_j( \mathit{X} \Psi_{2m})\xi_{km}^{(j)}.
\end{align*}
Finally, consider $ \mathit{X} P_{2k+1} - P_{2k+2}$. This has degree at most $2k+1$ and so may be expanded as follows:
\[
 \mathit{X} P_{2k+1} - P_{2k+2} = \sum_{m=0}^k \tilde{\mu}_{mk} P_{2m} + \sum_{m=0}^k \tilde{\lambda}_{mk} P_{2m+1}.
\]
Then after a calculation, we find
\begin{align*}
\tilde{\mu}_{mk} &= \delta_{k,m+1}c_k - \frac{h_k}{h_m} \sum_{j=1}^{D-1} \beta_j(\mathit{X} \Psi_{2m+1}) \eta_{km}^{(j)} - b_k \delta_{km} , \\
\tilde{\lambda}_{mk} &= \frac{h_k}{h_m} \sum_{j=1}^{D-1} \beta_j (\mathit{X} \Psi_{2m}) \eta_{km}^{(j)} + \tilde{a}_k \delta_{km}.
\end{align*}
Let us now derive a Christoffel--Darboux type formula for the pre-kernel \eqref{prekernel}. Note that the pre-kernel may be written
\begin{equation*}
S_n(x,y) = - \frac{1}{2} \gamma D {\rm e}^{-V(x) - V(y)} \sum_{k=0}^{n-1} \frac{P_{2k}(x) \Psi_{2k+1}(y) - P_{2k+1}(x) \Psi_{2k}(y)}{h_k}.
\end{equation*}
Then applying our recursion formulas, we find
\begin{align*}
(x-y) S_n(x,y) = - \frac{1}{2} \gamma D {\rm e}^{-V(x) - V(y)} \sum_{k=0}^{n-1} \left( \Circled{1}_k - \Circled{2}_k - \Circled{3}_k+\Circled{4}_k \right),
\end{align*}
where
\begin{align*}
\Circled{1}_k ={}& h_k^{-1} \left( P_{2k+1}(x) + \sum_{m=0}^k \mu_{mk} P_{2m}(x) + \sum_{m=0}^{k-1} \lambda_{mk} P_{2m+1}(x) \right) \Psi_{2k+1}(y) ,\\
\Circled{2}_k ={}& h_k^{-1} P_{2k}(x) \Bigg( \sum_{i=k}^{n-1} \sum_{j=1}^{D-1} \beta_j(\mathit{X} \Psi_{2k+1}) \big[ \eta_{ik}^{(j)} \Psi_{2i}(y) + \xi_{ik}^{(j)} \Psi_{2i+1}(y) \big] \\
& + \Psi_{2k+2}(y) + a_k \Psi_{2k+1}(y) + b_k \Psi_{2k}(y) + c_k \Psi_{2k-2}(y) + \sum_{j=1}^{D-1} \beta_j (\mathit{X} \Psi_{2k+1}) R^{(j)}_n(y) \Bigg), \\
\Circled{3}_k ={}& h_k^{-1} \left( P_{2k+2}(x) + \sum_{m=0}^k \tilde{\mu}_{mk} P_{2m}(x) + \sum_{m=0}^k \tilde{\lambda}_{mk} P_{2m+1}(x) \right) \Psi_{2k}(y), \\
\Circled{4}_k ={}& h_k^{-1} P_{2k+1}(x) \Bigg( \Psi_{2k+1}(y) + \tilde{a}_k \Psi_{2k}(y) + \sum_{j=1}^{D-1} \beta_j (\mathit{X} \Psi_{2k}) R_n^{(j)}(y) \\
& + \sum_{i=k}^{n-1} \sum_{j=1}^{D-1} \beta_j(\mathit{X} \Psi_{2k}) \big[ \eta_{ik}^{(j)} \Psi_{2i}(y) + \xi_{ik}^{(j)} \Psi_{2i+1}(y) \big] \Bigg).
\end{align*}
Let us then write
\begin{align*}
\sum_{k=0}^{n-1} \Circled{2}_k ={}& h_{n-1}^{-1} P_{2n-2}(x) \Psi_{2n}(y) + h_{n-1}^{-1} P_{2n}(x) \Psi_{2n-2}(y) \\
& + \sum_{j=1}^{D-1} \sum_{k=0}^{n-1} \beta_j (\mathit{X} \Psi_{2k+1}) h_k^{-1} P_{2k}(x) R^{(j)}_n(y) + \sum_{k=0}^{n-1} \Circled{5}_k,
\end{align*}
where
\begin{align*}
\Circled{5}_k ={}& h_{k-1}^{-1} P_{2k-2}(x) \Psi_{2k}(y) + h_k^{-1} P_{2k}(x) \left( a_k \Psi_{2k+1}(y) + b_k \Psi_{2k}(y)\right) - h_{k}^{-1} P_{2k+2}(x) \Psi_{2k}(y) \\
& + \sum_{m=0}^{k} h_m^{-1} P_{2m}(x) \sum_{j=1}^{D-1} \beta_j(\mathit{X} \Psi_{2m+1}) \big[ \eta_{km}^{(j)} \Psi_{2k}(y) + \xi_{km}^{(j)} \Psi_{2k+1}(y) \big],
\end{align*}
and similarly we may write
\begin{align*}
\sum_{k=0}^{n-1} \Circled{4}_k &= \sum_{j=1}^{D-1} \sum_{k=0}^{n-1} h_k^{-1} P_{2k+1}(x) \beta_j (\mathit{X} \Psi_{2k}) R_n^{(j)}(y) + \sum_{k=0}^{n-1} \Circled{6}_k,
\end{align*}
where
\begin{align*}
\Circled{6}_k ={}& h_k^{-1} P_{2k+1}(x) \Psi_{2k+1}(y) + \tilde{a}_k h_k^{-1} P_{2k+1}(x) \Psi_{2k}(y) \\
& + \sum_{m=0}^k h_m^{-1} P_{2m+1}(x) \sum_{j=1}^{D-1} \beta_j (\mathit{X} \Psi_{2m}) \big[ \eta^{(j)}_{km} \Psi_{2k}(y) + \ \xi^{(j)}_{km} \Psi_{2k+1}(y) \big].
\end{align*}
Thus
\begin{align*}
(x-y) S_n(x,y) ={}& \frac{\gamma D}{2} {\rm e}^{-V(x) - V(y)} \big( h_{n-1}^{-1} P_{2n-2}(x) \Psi_{2n}(y) + h_{n-1}^{-1} P_{2n}(x) \Psi_{2n-2}(y) \\
& + \sum_{j=1}^{D-1} \sum_{k=0}^{n-1} h_k^{-1} [ \beta_j (\mathit{X} \Psi_{2k+1}) P_{2k}(x) - \beta_j (\mathit{X} \Psi_{2k}) P_{2k+1}(x) ] R^{(j)}_n(y) \Bigg) \\
& + \frac{\gamma D}{2} {\rm e}^{-V(x) - V(y)} \sum_{k=0}^{n-1} \left( - \Circled{1}_k + \Circled{5}_k + \Circled{3}_k - \Circled{6}_k \right).
\end{align*}
The reader may verify that a miraculous cancellation occurs so that \[
- \Circled{1}_k + \Circled{5}_k + \Circled{3}_k - \Circled{6}_k = 0.
\]
 Finally, using our solutions of $A_n$ and \smash{$\widehat{A_n}$},
\begin{align*}
(x-y) S_n(x,y) ={}& \frac{\gamma D}{2} {\rm e}^{-V(x) - V(y)} \big( h_{n-1}^{-1} P_{2n-2}(x) \Psi_{2n}(y) + h_{n-1}^{-1} P_{2n}(x) \Psi_{2n-2}(y) \\
& + \sum_{j=1}^{D-1} \sum_{k=0}^{n-1} h_k^{-1} [ \beta_j (\mathit{X} \Psi_{2k+1}) P_{2k}(x) - \beta_j (\mathit{X} \Psi_{2k}) P_{2k+1}(x) ] R^{(j)}_n(y) \Bigg) \\
={}& - \frac{{\rm e}^{-V(x) - V(y)}}{4\pi {\rm i}} \big( A^{-1}_n(x) A_n(y) \big)_{21}.
\end{align*}
\begin{Remark}
We refer to this cancellation as \enquote{miraculous}, however one could have predicted that some sort of cancellation must occur simply from the form of the pre-kernel \eqref{prekernel}. Introduce the integral operator
\begin{align*}
\mathsf{S}_n \colon\ L^2(\mathbb{R}) \longrightarrow L^2(\mathbb{R}),\qquad
(\mathsf{S}_n \psi)(x) = \int_\mathbb{R} S_n(x,y) \psi(y) {\rm d}y,\qquad \psi \in L^2(\mathbb{R}).
\end{align*}
It is then easily seen that if $\{ H_k \}_{k \in \mathbb{N}}$ is the sequence of \textit{orthogonal} polynomials with respect to the weight $w = {\rm e}^{-2V}$, then $\mathsf{S}_n \big( H_k {\rm e}^{-V}\big) = \frac{1}{2} H_k {\rm e}^{-V}$ for $k \leq 2n-1$ and $\mathsf{S}_n \big( H_k {\rm e}^{-V}\big) = 0$ for $k \geq 2n+D-1$. It thus follows that
\begin{equation*}
\mathsf{S}_n = \frac{1}{2} \mathsf{K}_n + \sum_{j=2n}^{2n+D-2} \frac{1}{\nu_j} \mathsf{S}_n\big(H_j {\rm e}^{-V}\big) \otimes H_j {\rm e}^{-V},
\end{equation*}
where $\mathsf{K}_n = \sum_{k=0}^{n-1} \frac{1}{\nu_k} H_k \otimes H_k$ for $\nu_k = \int_\mathbb{R} H_k^2 {\rm e}^{-2V} {\rm d}x$. By the classical Christoffel--Darboux formula for orthogonal polynomials and the three term recurrence relation, it follows that ${(x-y) S_n(x,y)}$ has rank at most $D+1$. Indeed, this is one of the ideas behind Widom's method.
\end{Remark}

\section{A dynamical system}\label{dynamicsection}

This paper is part of a broader project to develop structures for $\beta = 1,4$ random matrix ensembles parallel to those of $\beta = 2$ ensembles. A central part of the $\beta = 2$ theory is the relation between orthogonal polynomials and the celebrated Toda lattice (see \cite[Chapter 2]{deiftbook}). One way to derive this relation is, starting with the Fokas--Its--Kitaev Riemann--Hilbert problem for orthogonal polynomials~\cite{fokas}, to pass to a problem with constant jump matrices and then by standard techniques of varying parameters one derives a Lax pair whose compatibility condition gives the Toda lattice in Flaschka variables. In this section we repeat this method to see what analogous dynamical system arises for our Riemann--Hilbert Problem~\ref{evenrhp}. The calculation is complicated and so we confine ourselves simply to deriving the compatibility condition, and leave the question of implications for the theory of integrable systems to future research.

We begin, analogously to the Toda lattice, by introducing a $t \in \mathbb{R}$ dependence into the potential as
\begin{equation}\label{deformation}
 V(z,t) := V_0(z) + tz.
\end{equation}
Let \smash{$\widehat{A_{n}}(\cdot,t)$} be the solution of Riemann--Hilbert Problem~\ref{dualeven} with potential $V(\cdot, t)$ and let \smash{$\widehat{A_{n}}(\cdot,t)$} have the following expansion at $z= \infty$:
\begin{gather*}
\widehat{A_{n}}(z,t) = \big( \mathbb{I}+ z^{-1} \widehat{A}_{n}^{(1)}(t) + z^{-2} \widehat{A}_{n}^{(2)}(t) + \mathcal{O}\big(z^{-3}\big) \big) \left(\begin{matrix}
z^{-2n-D+1} \\
& z^{2n} \\
& & z \\
& & & \ddots \\
& & & & z
\end{matrix} \right),\\ z \to \infty.
\end{gather*}
Introduce the variables \smash{$a_{ij}^{(1)}(n,t) \overset{\mathrm{def}}{=} \widehat{A}_{n}^{(1)}(t)_{ij}$} and \smash{$a_{ij}^{(2)}(n,t) \overset{\mathrm{def}}{=} \widehat{A}_{n}^{(2)}(t)_{ij}$} for $i,j \in [\![1, D+1]\!]$. In what follows, we shall often not write the dependence on $t$ since this is always understood. Furthermore, introduce the variable
\begin{align}\label{defb}
b_j(n,t) \overset{\mathrm{def}}{=} \sum_{k=3}^{D+1} a_{2k}^{(1)}(n,t) a_{kj}^{(1)}(n,t),\qquad j \in [\![3, D+1]\!] .
\end{align}
\begin{Theorem}\label{skewtoda} The matrix elements \smash{$a^{(1)}_{22}$}, \smash{$a^{(1)}_{2i}$}, \smash{$a^{(1)}_{i2}$}, \smash{$a^{(2)}_{22}$}, \smash{$a^{(2)}_{12}$}, \smash{$a^{(2)}_{i2}$} and $b_i$, for $i=3, \dots, D+1$, satisfy the following first-order dynamical system $(n \geq 0$, $t \in \mathbb{R})$:
\begin{gather*}
\frac{{\rm d} a_{22}^{(1)}(n) }{{\rm d}t} = \sum_{k=3}^{D+1} a_{2k}^{(1)}(n) a_{k2}^{(1)}(n) ,\qquad
\frac{{\rm d} a_{2i}^{(1)}(n)}{{\rm d}t} = b_i(n),\qquad i \in [\![3, D+1]\!], \\
\frac{{\rm d} a_{i2}^{(1)}(n)}{{\rm d}t} = a_{i2}^{(2)}(n) - a_{i2}^{(1)}(n) a_{22}^{(1)}(n),\qquad i \in [\![ 3, D+1]\!], \\
\frac{{\rm d} a_{22}^{(2)}(n)}{{\rm d}t} = \sum_{k=3}^{D+1} a_{2k}^{(1)}(n)a_{k2}^{(2)}(n), \qquad
 \frac{{\rm d}}{{\rm d}t} \ln a_{12}^{(2)}(n+1) =2a_{22}^{(1)}(n) -\sum_{j=3}^{D+1} a_{2j}^{(1)}(n)a_{j2}^{(2)}(n+1), \\
\frac{{\rm d}}{{\rm d}t}a_{i2}^{(2)}(n+1) = a_{i2}^{(2)}(n+1) a_{22}^{(1)}(n) + a_{i2}^{(1)}(n+1)a_{22}^{(2)}(n)+ a_{i2}^{(1)}(n) - a_{i2}^{(1)}(n+1) a_{22}^{(2)}(n+1) \\
\phantom{\frac{{\rm d}}{{\rm d}t}a_{i2}^{(2)}(n+1) =}{} + a_{i2}^{(1)}(n+2)\frac{ a_{12}^{(2)}(n+1)}{a_{12}^{(2)}(n+2)} - a_{i2}^{(1)}(n+1) a_{22}^{(1)}(n)^2 \\
\phantom{\frac{{\rm d}}{{\rm d}t}a_{i2}^{(2)}(n+1) =}{} - a_{i2}^{(1)}(n+1) \sum_{k=3}^{D+1} a_{2k}^{(1)}(n) a_{k2}^{(1)}(n),\qquad i \in [\![3, D+1]\!] ,
\end{gather*}
and for $n \geq 1$ and $t \in \mathbb{R}$,
\begin{align*}
\frac{{\rm d} b_i(n)}{{\rm d}t} ={}& \frac{a_{12}^{(2)}(n) }{a_{12}^{(2)}(n+1)}a_{2i}^{(1)}(n-1) + a_{22}^{(1)}(n+1) a_{22}^{(1)}(n) a_{2i}^{(1)}(n) + a_{2i}^{(1)}(n+1) \\
& -a_{22}^{(2)}(n+1)a_{2i}^{(1)}(n)+ a_{22}^{(2)}(n) a_{2i}^{(1)}(n) -a_{2i}^{(1)}(n) \big( a_{22}^{(1)}(n) \big)^2 - a_{22}^{(1)}(n) b_i(n)\\
& - 2 a_{2i}^{(1)}(n) \sum_{k=3}^{D+1}a_{2k}^{(1)}(n) a_{k2}^{(1)}(n) + a_{22}^{(1)}(n+1) b_i(n), \qquad i \in [\![3, D+1]\!].
\end{align*}
Note that \smash{$a_{12}^{(2)}(n+1) \neq 0$} so the above is well defined. The above dynamics are furthermore constrained by the relation
\begin{align}\label{identity}
\sum_{j=3}^{D+1} a_{2j}^{(1)}(n) a_{j2}^{(1)}(n+1) = 2.
\end{align}
Our \enquote{initial data} with respect to $n$ must satisfy $\smash{a_{22}^{(1)}(0,t) = a_{i2}^{(1)}(0,t) = a_{22}^{(2)}(0,t) = a_{12}^{(2)}(0,t)} =\smash{ a_{i2}^{(2)}(0,t) = b_i(0,t)=0}$ for $i \in [\![ 3, D+1]\!]$.
\end{Theorem}

The identity \eqref{identity} may be seen as analogous to \eqref{intriguing}. What follows in the rest of this section is a proof of Theorem \ref{skewtoda}.

\begin{Remark} Note that repeated differentiation of \eqref{identity} yields an infinite number of identities, so that the $k$-th derivative of \eqref{identity} yields an identity involving matrix elements at $k$ different values of $n$.
\end{Remark}

\begin{Remark}
Some of the matrix elements not included in Theorem \ref{skewtoda} may be related to those included by the following identities:
\begin{gather}
a_{i1}^{(1)}(n) = a_{21}^{(2)}(n) a_{i2}^{(1)}(n+1),\qquad i \in [\![ 3, D+1]\!], \label{deduce1} \\
a_{1i}^{(1)}(n+1) = a_{12}^{(2)}(n+1) a_{2i}^{(1)}(n),\qquad i \in [\![ 3, D+1]\!], \label{deduce2} \\
1 = a_{21}^{(2)}(n) a_{12}^{(2)}(n+1) ,\qquad
0 = a_{12}^{(1)}(n) = a_{21}^{(1)}(n) = a_{ij}^{(2)}(n),\nonumber\\ i \in [\![ 1, D+1]\!] ,\qquad j\in [\![ 3, D+1]\!] .\nonumber
\end{gather}
These identities can be seen directly from the solution, though \eqref{deduce1} and \eqref{deduce2} will be deduced in the course of proving Theorem \ref{skewtoda}. The matrix elements \smash{$a^{(2)}_{i1}$} for $i\neq 2$ are left undetermined by these relations. Furthermore, \smash{$a_{ij}^{(1)}(n,t)$}, for $i,j \geq 3$ do not enter into the dynamical system on their own but only in the combination of the variable $b_i(n,t)$ \eqref{defb}.
\end{Remark}

\begin{Example}\label{exactsolution} In the case $D=2$, it is possible to give an exact solution. Without loss of generality (by shifting and rescaling), one can take $V_0(z) = \frac{1}{2}z^2$. With this initial condition one finds the following:
\begin{align*}
\widehat{A}_{n}^{(1)}(t) &= \left( \begin{matrix}
-t (2n+1) & 0 & \frac{2\pi {\rm i} {\rm e}^{-\frac{1}{2}t^2}}{\sqrt{2}\left( n-\frac{1}{2} \right)!} \\[1mm]
0 & 2nt & - \frac{{\rm e}^{\frac{1}{2}t^2}n!}{\sqrt{2}} \\[1mm]
 \frac{{\rm e}^{\frac{1}{2}t^2} 2 \sqrt{2} \left( n + \frac{1}{2} \right)!}{2\pi {\rm i} } &- \frac{2\sqrt{2} {\rm e}^{-\frac{1}{2}t^2}}{(n-1)!} & t\end{matrix} \right), \\
\widehat{A}_{n}^{(2)}(t) &= \left( \begin{matrix}
 \ast & - {\rm i} \sqrt{\pi} \frac{2^{2n} {\rm e}^{-t^2}}{(2n-1)!} & 0 \\[1mm]
 -\frac{(2n+1)! \sqrt{\pi}}{2^{2n+1}} \frac{{\rm e}^{t^2}}{2\pi {\rm i}} & n(2n-1)\big(t^2 - \frac{1}{2}\big)+n & 0 \\[1mm]
\ast & - \frac{2\sqrt{2}(2n-1) t {\rm e}^{-\frac{1}{2}t^2}}{(n-1)!} & 0 \end{matrix} \right).
\end{align*}
We find also \smash{$b_3(n,t) = - \frac{t {\rm e}^{\frac{1}{2}t^2}n!}{\sqrt{2}}$}. The reader may verify that this is an exact solution of the dynamical system in Theorem \ref{skewtoda} for the case $D=2$.
\end{Example}

Let us introduce the $(D+1) \times (D+1)$ matrices
\begin{align*}
E_1 = \mathrm{diag}(1,\underbrace{0, \dots, 0}_{D \text{ times}}),\qquad
E_2 = \mathrm{diag}(0,1,\underbrace{0, \dots, 0}_{D-1 \text{ times}}) ,\qquad
\sigma = E_1 - E_2 .
\end{align*}
Let us then define
\begin{align*}
\widebreve{A}_{n}(z,t) \overset{\mathrm{def}}{=} \widehat{A_{n}}(z,t) {\rm e}^{ V(z,t) \sigma} .
\end{align*}
It is then easy to see that $\widebreve{A}_{n}$ satisfies the following Riemann--Hilbert problem.
\begin{rhp} Let $\Gamma$ be as in \eqref{gamma}. Then \smash{$\widebreve{A}_{n}$} has the following properties:
\begin{enumerate}\itemsep=0pt
\item[$(1)$] \smash{$z\mapsto \widebreve{A}_{n}(z,t)$} is an analytic function on $\mathbb{C}\setminus \Gamma$.
\item[$(2)$] \smash{$z\mapsto \widebreve{A}_{n}(z,t)$} has continuous left $(+)$ and right $(-)$ non-tangential boundary values$ x \mapsto \smash{\widebreve{A}_{n}^{\pm}(x,t)}$ for $x \in \Gamma \setminus \{ 0 \}$ which are related by the jump condition
\begin{align*}
\widebreve{A}_{n}^{+}(x,t) = \widebreve{A}_{n}^{-}(x,t) \left(
\begin{array}{ccccccccccccccccccc}
1 & & & & \\
-\chi_\mathbb{R} & 1 \\
-\chi_{\Gamma_1} & & 1 \\
\vdots & & & \ddots \\
-\chi_{\Gamma_{D-1}} & & & & 1 \\
\end{array} \right),\qquad x \in \Gamma \setminus \{ 0 \}.
\end{align*}
\item[$(3)$] For each $t \in \mathbb{R}$, the function \smash{$z \mapsto \widebreve{A}_{n}(z,t)$} is bounded in a neighbourhood of $0 \in \mathbb{C}$.
\item[$(4)$] For each $t \in \mathbb{R}$, \smash{$z \mapsto \widebreve{A}_{n}(z,t)$} satisfies the asymptotic normalisation
\begin{align*}
\widebreve{A}_{n}(z,t) = \big(\mathbb{I}+ \mathcal{O}\big(z^{-1}\big)\big) \left(\begin{matrix}
z^{-2n-D+1} \\
& z^{2n} \\
& & z \\
& & & \ddots \\
& & & & z
\end{matrix} \right) {\rm e}^{V(z,t) \sigma},\qquad z \to \infty.
\end{align*}
\end{enumerate}
\end{rhp}
Because the jump matrix is independent of both $n$ and $t$, standard Riemann--Hilbert theory now enables us to derive a dynamical system. In particular \smash{$\widebreve{A}_{n+1}(z,t)$} and \smash{$\frac{\partial \widebreve{A}_{n}(z,t)}{\partial t}$} both satisfy the first to third properties, differing only on the fourth property. Because of the continuity of the boundary values and boundedness at $0$, we find that
\begin{align*}
z \mapsto \widebreve{A}_{n+1}(z,t)\widebreve{A}_{n}(z,t)^{-1},\qquad
z \mapsto \frac{\partial \widebreve{A}_{n}(z,t)}{\partial t}
\widebreve{A}_{n}(z,t)^{-1}
\end{align*}
extend to entire functions on $\mathbb{C}$. Let us write the asymptotic expansion ($z \to \infty$)
\begin{align*}
\widebreve{A}_{n}(z,t) = \big(\mathbb{I}+ z^{-1} \widehat{A}_{n}^{(1)}(t) + z^{-2} \widehat{A}_{n}^{(2)}(t) + \mathcal{O}\big(z^{-3}\big) \big) \left(\begin{matrix}
z^{-2n-D+1} \\
& z^{2n} \\
& & z \\
& & & \ddots \\
& & & & z
\end{matrix} \right) {\rm e}^{ V(z,t) \sigma}.
\end{align*}
A short calculation yields the \textit{Lax pair},
\begin{gather}
\widebreve{A}_{n+1}(z,t) = \big( z^2 E_2 + z \rho_n^{(1)}(t) + \rho_n^{(2)}(t) \big) \widebreve{A}_{n}(z,t), \label{Laxpair1} \\
\frac{\partial \widebreve{A}_{n}(z,t)}{\partial t} = ( z \sigma + \kappa_n(t) ) \widebreve{A}_{n}(z,t), \label{Laxpair2}
\end{gather}
where
\begin{gather*}
\rho_n^{(1)}(t) = \widehat{A}_{n+1}^{(1)}(t) E_2 - E_2 \widehat{A}_{n}^{(1)}(t), \\
\rho_n^{(2)}(t) = \mathbb{I} - E_1 - E_2 + \widehat{A}_{n+1}^{(2)}(t) E_2 + E_2 \big( \widehat{A}_{n}^{(1)}(t)^2 - \widehat{A}_{n}^{(2)}(t) \big) - \widehat{A}_{n+1}^{(1)}(t) E_2 \widehat{A}_{n}^{(1)}(t), \\
\kappa_n(t) = [\widehat{A}_{n}^{(1)}(t),\sigma].
\end{gather*}
\begin{Remark} The reader will observe that the jump matrices of \smash{$\widebreve{A}_n$} are not only independent of $n$ and $t$ but also independent of $z$. Thus one could also derive $(n,z)$ and $(t,z)$ Lax pairs. Indeed, one could consider more general deformations than \eqref{deformation}.
\end{Remark}

\begin{Remark}[analogue of three term recurrence]\label{recurrence} We note that \eqref{Laxpair1} implies a similar equation for \smash{$\widehat{A_n}$}, i.e.,
\begin{equation*}
\widehat{A_{n+1}}(z) = \big( z^2 E_2 + z \rho_n^{(1)} + \rho_n^{(2)} \big) \widehat{A_{n}}(z) .
\end{equation*}
Considering the second column of both sides yields a recursion relation, quadratic in $z$, which may be regarded as a $\beta = 4$ analogue of the three term recurrence for orthogonal polynomials (recall equations \eqref{12Ahat}, \eqref{22Ahat}, \eqref{i2Ahat}).
\end{Remark}

Before proceeding, we must indicate a point of rigour. To compute this Lax pair we have exchanged a $z \to \infty$ limit with a derivative with respect to $t$. How can this be justified? The proof is elementary but tedious to write out and so we only sketch how this is done. The solution of our even problem \smash{$\widehat{A_n}$} is expressible as rational combinations of integrals of the form
\[
\int_{\Gamma_j} x^k {\rm e}^{-V(x,t)} {\rm d}x, \qquad \int_{\mathbb{R}} x^k {\rm e}^{-2V(x,t)} {\rm d}x, \qquad \int_{\Gamma_j} \frac{x^k}{x-z} {\rm e}^{-V(x,t)} {\rm d}x, \qquad \int_{\mathbb{R}} \frac{x^k}{x-z} {\rm e}^{-2V(x,t)} {\rm d}x.
\]
 The proof reduces to showing that such integrals are $\mathscr{C}^\infty$ functions of $t$ and that we can differentiate under the integral sign. Both can be justified by a dominated convergence argument. This implies that~\smash{$\widehat{A}_n^{(1)}$} and \smash{$\widehat{A}_n^{(2)}$} are $\mathscr{C}^\infty$ functions of $t$.

Moving forward, let us write
\begin{align*}
\rho(z,t,n) = z^2 E_2 + z \rho_n^{(1)}(t) + \rho_n^{(2)}(t),\qquad
\lambda(z,t,n) = z \sigma + \kappa_n(t) .
\end{align*}
Then the compatibility condition (or \textit{zero curvature condition}) of the above Lax pair~\eqref{Laxpair1} and~\eqref{Laxpair2} is
\begin{align}\label{compatibility}
\frac{\partial \rho(z,t,n)}{\partial t} + \rho(z,t,n) \lambda(z,t,n)-\lambda(z,t,n+1) \rho(z,t,n)=0, \qquad \forall z \in \mathbb{C} .
\end{align}
The left-hand side is at most a cubic polynomial in $z$, all of whose coefficients must vanish. In fact the coefficient of $z^3$ is $[E_2, \sigma] = 0$ which yields no information. The coefficient of $z^2$ yields the equation
\begin{align*}
E_2 \kappa_n(t) - \kappa_{n+1}(t) E_2 + \big[ \rho_n^{(1)}(t),\sigma\big] = 0.
\end{align*}
However, when we substitute in our formulas for \smash{$\rho_n^{(1)}$} and $\kappa_n$ we find that the left-hand side is identically zero. So the quadratic coefficient also yields no information. Thus the left-hand side of \eqref{compatibility} is in fact a linear polynomial in $z$.

The coefficient of $z$ in \eqref{compatibility} yields
\begin{align}\label{linear}
\frac{{\rm d}\rho_n^{(1)}(t)}{{\rm d}t} + \rho_n^{(1)}(t) \kappa_n(t) - \kappa_{n+1}(t) \rho_n^{(1)}(t) + \big[\rho_n^{(2)}(t),\sigma\big] = 0,
\end{align}
while the constant term gives
\begin{align}\label{constant}
\frac{{\rm d} \rho_n^{(2)}(t)}{{\rm d}t}+\rho_n^{(2)}(t)\kappa_n(t)-\kappa_{n+1}(t) \rho_n^{(2)}(t)=0.
\end{align}
Equation \eqref{linear}, after substituting our expressions for \smash{$\rho^{(1)}_{n}(t)$} and \smash{$\rho^{(2)}_n(t)$}, yields
\begin{gather}
-\widehat{A}^{(2)}_{n+1} E_2 + \widehat{A}^{(1)}_{n+1} E_2 \widehat{A}^{(1)}_{n+1} E_2 - E_1 \widehat{A}^{(2)}_{n+1} E_2 - E_2 \widehat{A}^{(2)}_n + E_2 \widehat{A}^{(2)}_n E_2 - E_2 \widehat{A}^{(2)}_n E_1+E_2 \widehat{A}^{(2)}_{n+1} E_2\nonumber \\
\qquad{} - E_2 \frac{{\rm d} \widehat{A}^{(1)}_n}{{\rm d}t}\! - E_2 \big(\widehat{A}^{(1)}_{n + 1}\big)^2 E_2
-E_2 \widehat{A}^{(1)}_{n} E_2 \widehat{A}^{(1)}_n + E_2 \big(\widehat{A}^{(1)}_n\big)^2\! + \frac{{\rm d} \widehat{A}^{(1)}_{n+1}}{{\rm d}t} E_2 + E_1 \big(\widehat{A}^{(1)}_{n+1}\big)^2 E_2\nonumber\\
\quad\qquad= 0.\label{linearexpansion}
\end{gather}
The $(1,1)$ matrix element of \eqref{linearexpansion} is $0$. The $(1,2)$ matrix element yields the identity
\begin{equation}\label{identity1}
2 a_{12}^{(2)} (n) =\sum_{k=3}^{D+1} a_{1k}^{(1)} (n) a_{k2}^{(1)}(n).
\end{equation}
The $(2,1)$ matrix element yields
\begin{equation}\label{identity2}
2 a_{21}^{(2)} (n) =\sum_{k=3}^{D+1} a_{2k}^{(1)} (n) a_{k1}^{(1)}(n).
\end{equation}
The $(2,2)$ element gives that the quantity
\begin{equation*}
(n,t) \mapsto \frac{\rm d}{{\rm d}t}a_{22}^{(1)}(n,t) - \sum_{k=3}^{D+1} a_{2k}^{(1)}(n,t) a_{k2}^{(1)}(n,t)
\end{equation*}
is independent of $n$.
\begin{Lemma}\label{nequal0}
\begin{equation*}
\frac{\rm d}{{\rm d}t}a_{22}^{(1)}(n) = \sum_{k=3}^{D+1} a_{2k}^{(1)}(n) a_{k2}^{(1)}(n).
\end{equation*}
\end{Lemma}
\begin{proof}
To show this, we need only show that
\begin{equation*}
\frac{\rm d}{{\rm d}t}a_{22}^{(1)}(0,t) - \sum_{k=3}^{D+1} a_{2k}^{(1)}(0,t) a_{k2}^{(1)}(0,t) = 0.
\end{equation*}
However, if $n=0$, then $\big(\widehat{A_n}\big)_{22} = P_{2n} = P_0 = 1$, and hence \smash{$a_{22}^{(1)}(0,t) = 0$}. Similarly, $\big(\widehat{A_n}\big)_{k2} = 0$ for $k \geq 3$, and so \smash{$a_{k2}^{(1)}(0,t) = 0$}.
\end{proof}

The $(1,i)$ and $(i,1)$ matrix elements for $i\geq 3$ give $0$. The $(2,i)$ matrix element for $i \geq 3$ gives
\begin{align*}
\frac{{\rm d} a_{2i}^{(1)}(n)}{{\rm d}t} = \sum_{k=3}^{D+1} a_{2k}^{(1)}(n) a_{ki}^{(1)}(n), \qquad i \in [\![ 3, D+1]\!].
\end{align*}
Finally, the $(i,2)$ matrix element for $i\geq 3$ gives the equation
\begin{align*}
\frac{{\rm d} a_{i2}^{(1)}(n,t)}{{\rm d}t} = a_{i2}^{(2)}(n,t) - a_{i2}^{(1)}(n,t) a_{22}^{(1)}(n,t), \qquad i \in [\![ 3, D+1]\!].
\end{align*}
Next let us move onto the constant term \eqref{constant}, which gives us the complicated expression
\begin{gather}
 E_2 \widehat{A}^{(2)}_n E_1 \widehat{A}^{(1)}_n - E_2 \widehat{A}^{(2)}_n \widehat{A}^{(1)}_n E_1 + E_2 \widehat{A}^{(2)}_n \widehat{A}^{(1)}_n E_2 - E_2 \widehat{A}^{(1)}_{n+1} \widehat{A}^{(2)}_{n+1} E_2
 + E_2 \widehat{A}^{(1)}_{n+1} E_2\nonumber
 \\
\qquad{}- E_2 \big(\widehat{A}^{(1)}_n\big)^3 E_2
- \widehat{A}^{(1)}_{n+1} E_2 \widehat{A}^{(1)}_n E_2 \widehat{A}^{(1)}_n - \widehat{A}^{(2)}_{n+1} E_2 \widehat{A}^{(1)}_n E_2
 + E_2 \widehat{A}^{(1)}_{n+1} E_2 \widehat{A}^{(2)}_n\nonumber\\
 \qquad{}+ E_1 \widehat{A}^{(1)}_{n+1} \widehat{A}^{(2)}_{n+1} E_2
 - \widehat{A}^{(1)}_{n+1} E_2 \big(\widehat{A}^{(1)}_n\big)^2 E_1 + E_2 \widehat{A}^{(1)}_n \frac{{\rm d} \widehat{A}^{(1)}_n }{{\rm d}t}
- E_1 \big(\widehat{A}^{(1)}_{n+1}\big)^2 E_2 \widehat{A}^{(1)}_n\nonumber \\
 \qquad{}- E_2 \frac{ {\rm d} \widehat{A}^{(2)}_n}{{\rm d}t}
- E_2 \widehat{A}^{(1)}_{n+1} E_2 \big(\widehat{A}^{(1)}_n \big)^2 - E_1 \widehat{A}^{(1)}_n E_1 + E_2 \frac{ {\rm d} \widehat{A}^{(1)}_n}{{\rm d}t} \widehat{A}^{(1)}_n
 - \widehat{A}^{(1)}_{n+1} E_2 \frac{ {\rm d} \widehat{A}^{(1)}_n}{{\rm d}t}\nonumber\\
\qquad{}+ \frac{ {\rm d} \widehat{A}^{(2)}_{n+1}}{{\rm d}t} E_2 - E_2 \widehat{A}^{(1)}_{n+1}
- E_2 \big(\widehat{A}^{(1)}_n \big)^2 E_1 \widehat{A}^{(1)}_n
 + E_2 \big(\widehat{A}^{(1)}_{n+1}\big)^2 E_2 \widehat{A}^{(1)}_n + \widehat{A}^{(2)}_{n+1} E_2 \widehat{A}^{(1)}_n \nonumber\\
\qquad{}+ \widehat{A}^{(1)}_n E_1 - \widehat{A}^{(1)}_{n+1} E_2 \widehat{A}^{(2)}_n
 - \widehat{A}^{(1)}_n E_2 + E_2 \widehat{A}^{(1)}_n E_2
+ E_2 \big(\widehat{A}^{(1)}_n\big)^2 E_2 \widehat{A}^{(1)}_n \nonumber \\
\qquad{}- \widehat{A}^{(1)}_{n+1} E_2 \widehat{A}^{(1)}_{n+1} E_2 \widehat{A}^{(1)}_n
 - \frac{{\rm d} \widehat{A}^{(1)}_{n+1}}{{\rm d}t} E_2 \widehat{A}^{(1)}_n + \widehat{A}^{(1)}_{n+1} E_2 \widehat{A}^{(2)}_{n+1} E_2
+ E_2 \big(\widehat{A}^{(1)}_n\big)^3 E_1 + E_1 \widehat{A}^{(1)}_{n+1} \nonumber\\
\qquad{}+ \widehat{A}^{(1)}_{n+1} E_2 \big(\widehat{A}^{(1)}_n\big)^2 E_2
+ \widehat{A}^{(1)}_{n+1} E_2 \big(\widehat{A}^{(1)}_n\big)^2
 -E_2 \widehat{A}^{(2)}_n E_2 \widehat{A}^{(1)}_n - E_1 \widehat{A}^{(1)}_{n+1} E_1 \nonumber\\
 \qquad{}- \widehat{A}^{(1)}_{n+1} E_1 \widehat{A}^{(2)}_{n+1} E_2 = 0.\label{constant2}
\end{gather}
The $(1,1)$ matrix element of \eqref{constant2} is $0$. The $(2,2)$ matrix element yields, after some simplification, the relation
\begin{align*}
\frac{{\rm d}a_{22}^{(2)}(n+1)}{{\rm d}t} - \sum_{k=3}^{D+1} a_{2k}^{(1)}(n+1)a_{k2}^{(2)}(n+1) = \frac{{\rm d} a_{22}^{(2)}(n)}{{\rm d}t} - \sum_{k=3}^{D+1} a_{2k}^{(1)}(n)a_{k2}^{(2)}(n).
\end{align*}
\begin{Lemma}
\begin{equation*}
\frac{{\rm d} a_{22}^{(2)}(n)}{{\rm d}t} = \sum_{k=3}^{D+1} a_{2k}^{(1)}(n)a_{k2}^{(2)}(n).
\end{equation*}
\end{Lemma}
\begin{proof}
Follows by the same observations as in Lemma \ref{nequal0}.
\end{proof}

The $(2,1)$ matrix element of \eqref{constant2} gives the following equation:
\begin{equation}\label{constant21}
 2 \big(a_{22}^{(1)}(n) - a_{22}^{(1)}(n+1) \big)a_{21}^{(2)}(n) + \frac{\rm d}{{\rm d}t} a_{21}^{(2)}(n)
+ \sum_{j=3}^{D+1} b_j(n) a_{j1}^{(1)}(n) = 0.
\end{equation}
The $(1,2)$ matrix element gives
\begin{equation}\label{constant12}
 \frac{\rm d}{{\rm d}t} a_{12}^{(2)}(n+1)+\sum_{k=3}^{D+1} a_{1k}^{(1)}(n+1)a_{k2}^{(2)}(n+1) - 2 a_{22}^{(1)}(n) a_{12}^{(2)}(n+1) = 0.
\end{equation}
The $(i,1)$ matrix element for $i\geq 3$ gives
\begin{align}\label{identity3}
a_{i1}^{(1)}(n) = a_{21}^{(2)}(n) a_{i2}^{(1)}(n+1),\qquad i \in [\![ 3, D+1]\!].
\end{align}
The $(1,i)$ matrix element gives
\begin{align}\label{identity4}
a_{1i}^{(1)}(n+1) = a_{12}^{(2)}(n+1) a_{2i}^{(1)}(n), \qquad i \in [\![ 3, D+1]\!].
\end{align}
The $(i,2)$ matrix element yields after some simplification
\begin{align*}
\frac{\rm d}{{\rm d}t}a_{i2}^{(2)}(n+1) ={}& a_{i2}^{(2)}(n+1) a_{22}^{(1)}(n) + a_{i2}^{(1)}(n+1)a_{22}^{(2)}(n) + a_{i2}^{(1)}(n)- a_{i2}^{(1)}(n+1) a_{22}^{(2)}(n+1) \\
& - a_{i2}^{(1)}(n+1) \sum_{k=3}^{D+1} a_{2k}^{(1)}(n) a_{k2}^{(1)}(n)+ a_{21}^{(2)}(n+1) a_{i2}^{(1)}(n+2) a_{12}^{(2)}(n+1) \\
& - a_{i2}^{(1)}(n+1) a_{22}^{(1)}(n)^2, \qquad i \in [\![ 3, D+1]\!] .
\end{align*}
The $(2,i)$ matrix element gives
\begin{gather*}
 -a^{(2)}_{21}(n) a_{12}^{(2)}(n) a_{2i}^{(1)}(n-1) - a_{22}^{(1)}(n+1) a_{22}^{(1)}(n) a_{2i}^{(1)}(n) - a_{2i}^{(1)}(n+1) +a_{22}^{(2)}(n+1)a_{2i}^{(1)}(n) \\
\qquad{}+a_{2i}^{(1)}(n) \big( a_{22}^{(1)}(n) \big)^2 + 2 a_{2i}^{(1)}(n) \sum_{k=3}^{D+1}a_{2k}^{(1)}(n) a_{k2}^{(1)}(n) + \frac{{\rm d} b_i(n)}{{\rm d}t} \\
\qquad{}+ \big(a_{22}^{(1)}(n) - a_{22}^{(1)}(n+1)\big) b_i(n) - a_{22}^{(2)}(n) a_{2i}^{(1)}(n) = 0,\qquad i \in [\![ 3, D+1]\!].
\end{gather*}
Let us now combine our formulas. Substituting \eqref{identity3} into \eqref{identity2}, we find
\[
\sum_{k=3}^{D+1} a_{2k}^{(1)}(n) a_{k2}^{(1)}(n+1) = 2.
\]
We would have found the same if we had substituted \eqref{identity4} into \eqref{identity1}.
Note that if we combine~\eqref{constant21} and \eqref{constant12}, we find
\begin{align*}
 &\frac{\rm d}{{\rm d}t} \ln a_{21}^{(2)}(n) + \frac{\rm d}{{\rm d}t} \ln a_{12}^{(2)}(n+1) \\
& \qquad = 2 a_{22}^{(1)}(n+1) - \sum_{k=3}^{D+1} b_k(n) a_{k2}^{(1)}(n+1)
- \sum_{k=3}^{D+1} a_{2k}^{(1)}(n) a_{k2}^{(2)}(n+1) \\
&\qquad = - \frac{\rm d}{{\rm d}t}\underbrace{ \sum_{k=3}^{D+1} a_{2k}^{(1)}(n) a_{k2}^{(1)}(n+1)}_{=2} = 0,
\end{align*}
which we already knew since \smash{$a_{21}^{(2)}(n) a_{12}^{(2)}(n+1) = 1$}. Finally, the $(i,j)$ matrix element of \eqref{constant2} for $i,j \geq 3$, after substituting the identities we have derived so far, is identically zero so yields no further information. This completes the proof of Theorem \ref{skewtoda}.

\subsection*{Acknowledgements}

The author would like to express his gratitude to Thomas Bothner and Mattia Cafasso, whose comments and suggestions have been invaluable for this paper. The author would also like to express his gratitude to the anonymous referees whose comments have improved the presentation of the paper and have drawn attention to connections to recent work. This work was supported by the UK Engineering and Physical Sciences Research Council through grant EP/T013893/2 and by the European Research Council (ERC) under the European Union Horizon 2020 research and innovation program (grant agreement No. 884584).

\pdfbookmark[1]{References}{ref}
\LastPageEnding

\end{document}